\documentclass[
  aps,
  prc,
  superscriptaddress,
  unsortedaddress,
  twocolumn,
  preprintnumbers
]{revtex4}

\usepackage{amsmath,amssymb}
\usepackage{physics}
\usepackage{graphicx}
\usepackage{xcolor}

\usepackage{ulem}
\usepackage{bm}
\newcommand{\txr}[1]{\textrm{#1}}
\newcommand{\trL}{\txr{L}}
\newcommand{\vK}{\vb*{K}}
\newcommand{\vR}{\vb*{R}}
\newcommand{\vkap}{\vb*{\kappa}}
\newcommand{\nn}{\nonumber\\}

\begin{document}

\title{Importance of the deuteron breakup in the deuteron knockout reaction}

\author{Yoshiki~Chazono}
\email[]{yoshiki.chazono@riken.jp}
\affiliation{Research Center for Nuclear Physics,
Ibaraki, Osaka 567-0047, Japan}
\affiliation{RIKEN Nishina Center for Accelerator-Based Science,
2-1 Hirosawa, Wako 351-0198, Japan}

\author{Kazuki~Yoshida}
\affiliation{Advanced Science Research Center, Japan Atomic Energy Agency,
Tokai, Ibaraki 319-1195, Japan}

\author{Kazuyuki~Ogata}
\affiliation{Department of Physics, Kyushu University,
Fukuoka 819-0395, Japan}
\affiliation{Research Center for Nuclear Physics,
Ibaraki, Osaka 567-0047, Japan}

\date{\today}

\begin{abstract}
\noindent
\textbf{Background:} An isoscalar $pn$ pair is expected to emerge in nuclei that have similar proton and neutron numbers and it may be a candidate for a deuteron ``cluster.'' There is, however, no experimental evidence for it.\\
\textbf{Purpose:} The purpose of this paper is to construct a new reaction model for the ($p,pd$) reaction including the deuteron breakup in the elementary process and the deuteron reformation by the final-state interactions (FSIs). How these processes contribute to the observables of the reaction is investigated.\\
\textbf{Methods:} The distorted wave impulse approximation is extended in twofold. The elementary processes of the ($p,pd$), i.e., the $p$-$d$ elastic scattering and $d(p,p)pn$ reaction, are described with an impulse picture employing a nucleon-nucleon effective interaction. The three-body scattering waves in the final state of the ($p,pd$) reaction are calculated with the continuum-discretized coupled-channels method. The triple differential cross section (TDX) of the ($p,pd$) reaction is calculated with the new model.\\
\textbf{Results:} The elementary processes are described reasonably well with the present model. As for the ($p,pd$) reaction, the deuteron reformation can either increase or decrease the TDX height depending on the interference between the elastic and breakup channel of deuteron, while the \textit{back-coupling} effect always decreases it.\\
\textbf{Conclusions:} It is shown that the deuteron reformation significantly changes the TDX of the ($p,pd$) reaction through the interference. It is important to include this process to quantitatively discuss the ($p,pd$) cross sections in view of the deuteron formation in nuclei. For more quantitative discussion regarding the experimental data, further improvement will be necessary.
\end{abstract}


\maketitle

\section{Introduction \label{sec1}}
From the analysis of many experimental data, it has been found that the ground and low-lying excited states of stable nuclei can be described with the independent-particle picture accounting for about $65$~\% \cite{JJKelly96, GJKramer01, TWakasa17}. This indicates that the remaining $35$~\% is understood as a result of some correlations between nucleons, which cannot be described by the single-particle model. An isovector and spin-singlet pairing correlation of two protons or neutrons \cite{GFBertsch13, DMBrink05} is one of the most well-known nucleon correlations. This correlation has been intensively studied for many years and has been found to play an important role in explaining properties such as moment of inertia and even-odd staggering in the binding energy of nuclei. A short-range tensor correlation, which is thought to introduce a high-momentum component to the relative motion of proton-neutron pairs in nuclei, has also been a hot subject in nuclear physics \cite{KSEgiyan03, KSEgiyan06, RSubedi08, RShneor07, MMSargsian05, RSchiavilla07}.

Another type of nucleon correlation is a spatial correlation. The alpha ($\alpha$) cluster structure of atomic nuclei is known as a typical example, and the deuteron ($d$) ``cluster'' may exist in nuclei. The proton-neutron ($pn$) pair is thought to be formed in the $N \approx Z$ nuclei, in which $N$ and $Z$ are the neutron and proton numbers of nuclei, because the shell structures of protons and neutrons are close to each other around the Fermi surface and the overlap of wave functions is large \cite{ALGoodman98, SFrauendorf14}. It has been suggested by the nuclear energy-density functional (EDF) approach that the isoscalar $pn$ pairing vibrational mode possibly emerges in $N=Z$ nuclei \cite{KenichiYoshida14, ELitvinova18}. The cluster orbital shell model (COSM) approach in which $^{18}$F ($N=Z=9$) is treated as a three-body system of $^{16}\txr{O}+p+n$ also suggests the formation of a deuteron-like $pn$ pair \cite{HMasui16}. The understanding of the $pn$ pair is getting more important since the invention of the unstable beams, which has provided more opportunities to study $N \approx Z$ nuclei in medium- and heavy-mass regions. The analysis of experimental data in the past, however, has not shown clear evidence for the existence of the deuteron ``cluster'' or deuteron-like $pn$ pair in nuclei \cite{SFrauendorf14}.

In the present study, the proton-induced deuteron knockout ($p,pd$) reaction is discussed to overcome the situation. In the experiment performed 40 years ago at the University of Maryland, it was reported that the cross section of the $^{16}$O($p,pd$)$^{14}$N \cite{CSamanta82} is almost half that of the $^{16}$O($p,2p$)$^{15}$N \cite{CSamanta86}. This result may indicate that the existence probability of deuteron in $^{16}$O is surprisingly high and hence the ($p,pd$) reaction can be a good probe for the $pn$ correlations including the deuteron ``cluster.'' To describe this reaction, it is important to treat the fragility of the deuteron properly. The deuteron can be easily broken up by the incident proton in the elementary process. In addition, the knocked-out deuteron is expected to go through transition between the bound and breakup states by the final-state interactions (FSIs). Furthermore, the deuteron broken up in the elementary process can reform a deuteron by the FSIs. These processes are not included in the distorted wave impulse approximation (DWIA) framework \cite{GJacob66, GJacob73, NSChant77, NSChant83, TWakasa17}, which is the standard reaction model for describing the knockout reactions as employed in the ($p,pd$) analysis of Ref.~\cite{CSamanta82}. Therefore, even if measurement results of deuteron knockout reactions are systematically obtained, it is not possible to conclude clearly whether deuterons exist in nuclei or not by the DWIA analysis. Thus, a reaction model beyond DWIA is necessary. The purposes of this paper are to construct such a reaction model and to investigate how the deuteron breakup in the elementary process and deuteron reformation by the FSIs contribute to the observables of the ($p,pd$) reaction. To achieve this, DWIA and the continuum-discretized coupled-channels method (CDCC) \cite{MKamimura86, NAustern87, MYahiro12} are combined in this work. We call this reaction model \textit{CDCCIA}.

It is worth mentioning that in hadron physics, the combination of DWIA and CDCC was used in the description of a kaon-induced $^3_\Lambda$He-production reaction, i.e., $^3$He($K^-, \pi^-$)$^3_\Lambda$He reaction \cite{THarada15}. In that study, DWIA was adopted to describe the elementary process $K^- n \rightarrow \Lambda \pi^-$, and CDCC was used for describing the decay process $^3_\Lambda$He$\rightarrow p + p + \Lambda$. Because $^3_\Lambda$He has no bound state and continuum states of the three-body system inevitably appear, CDCC was employed as a method to deal with it. On the other hand, in our study, by incorporating deuteron breakup $d(p,p)pn$ in the elementary process of the ($p,pd$) reaction, a new degree of freedom, i.e., deuteron reformations by FSIs, is activated. Therefore, CDCCIA in the present article can describe reaction processes that were not taken into account in the DWIA + CDCC study in Ref.~\cite{THarada15}.

The construction of this article is as follows. In Section~\ref{sec2}, we introduce the reaction model to describe the ($p,pd$) reaction including the deuteron breakup in the elementary processes and deuteron reformation by the FSIs. The numerical results of cross sections of the elementary processes, i.e., $p$-$d$ elastic scattering and $d(p,p)pn$ reaction, and those of the ($p,pd$) reactions on $^{16}$O, $^{40}$Ca, and $^{56}$Ni targets are shown in Sec.~\ref{sec3}. Finally a summary and perspective are given in Sec.~\ref{sec4}.

\section{Theoretical framework \label{sec2}}
The proton-induced deuteron knockout ($p,pd$) reaction in normal kinematics is considered. The incident and emitted protons are labeled as particles 0 and 1, respectively, and the knocked-out deuteron is referred to as particle 2. The target (residual) nucleus is labeled as A (B), whose mass number is denoted by $A$ ($B$). All quantities are evaluated in the center-of-mass (c.m.)~frame of the $p$-A system, and the superscript L is attached to those evaluated in the laboratory (L) frame. The spin-orbit part in each optical potential is ignored. In the formulation, the Coulomb interaction is not discussed explicitly for simplicity. The reaction is assumed to be coplanar, i.e., all momenta of the particles are on the $z$-$x$ plane, with taking the $z$-axis to be parallel to the direction of the incident beam.

\begin{figure}[htbp]
\begin{center}
\includegraphics[width=0.39\textwidth]{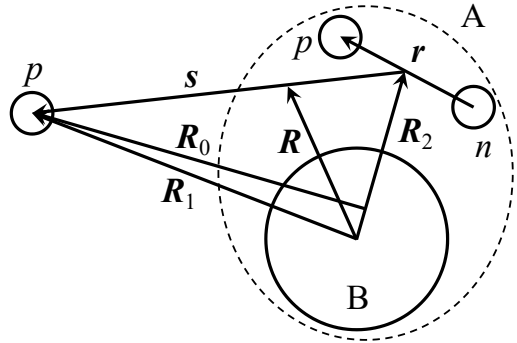}
\caption{Spatial coordinates of the A($p,pd$)B reaction system. \label{fig1}}
\end{center}
\end{figure}

The spatial coordinates to describe the A($p,pd$)B reaction are shown in Fig.~\ref{fig1}. The coordinate of particle 0 (1) concerning the c.m.~of A (B) is denoted by $\vR_0$ ($\vR_1$), and that of the c.m.~of particle 2 for B is written by $\vR_2$. One finds from Fig.~\ref{fig1} that $\vR_2$ also represents the relative coordinate between B and the $pn$ pair in the initial state. The coordinate between the proton and neutron in the deuteron is represented by $\vb*{r}$. The relative coordinate between the proton and the c.m.~of the deuteron is denoted by $\vb*{s}$.

In CDCCIA, the transition matrix of the ($p,pd$) reaction is given by
\begin{align}
T =
&\bra{\chi^{(-)}_{1, \vK_1} (\vR_1)
\Psi^{(-)}_{p n \txr{B}, \vK_2} (\vb*{r}, \vR_2)}
[t_{1p} (\vb*{s}_{1p}) \nn
&\quad + t_{1n} (\vb*{s}_{1n})]
\hat{\mathcal{A}}_\txr{elm}
\ket{\chi^{(+)}_{0, \vK_0} (\vR_0)
\varphi_d (\vR_2) \Phi_d (\vb*{r})},
\label{eq1}
\end{align}
in which $\chi_{0, \vK_0}$ and $\chi_{1, \vK_1}$ are the distorted waves of the $p$-A and $p$-B systems, respectively. $\vK_i$ ($i=0$, 1, or 2) is the momentum (in unit of $\hbar$) of particle $i$ in the asymptotic region. The $d$-B system in the final state is treated as the three-body system consisting of the proton, neutron, and B; $\Psi_{p n \txr{B}, \vK_2}$ is the three-body scattering wave of this system described with CDCC. The description of $\Psi_{p n \txr{B}, \vK_2}$ is given below. The wave functions with the superscripts $(+)$ and $(-)$ satisfy the outgoing and incoming boundary conditions, respectively. $\varphi_d$ is the bound-state wave function of the $d$-B system and $\Phi_d$ is the internal wave function of the deuteron. $t_{1p}$ and $t_{1n}$ are, respectively, the $p$-$p$ and $p$-$n$ effective interactions in free space with
\begin{align}
\vb*{s}_{1p} \equiv - \vb*{s} + \frac{1}{2} \vb*{r}, \quad
\vb*{s}_{1n} \equiv \vb*{s} + \frac{1}{2} \vb*{r}.
\end{align}
$\hat{\mathcal{A}}_\txr{elm}$ is the antisymmetrization operator defined by
\begin{align}
\hat{\mathcal{A}}_\txr{elm} \equiv
\frac{1}{\sqrt{3}} \qty[1 - \hat{P}_{1p} - \hat{P}_{1n}],
\end{align}
in which $\hat{P}_{1p}$ ($\hat{P}_{1n}$) exchanges particle 1 and the proton (neutron) in the deuteron. The normalization coefficient $1 / \sqrt{3}$ comes from the antisymmetrization of three nucleons when an antisymmetrized internal wave function $\Phi_d$ of the $pn$ system is adopted.

$\Psi_{p n \txr{B}, \vK_2}$ satisfies the following Schr\"{o}dinger equation
\begin{align}
\qty(\hat{H}^\dagger_{p n \txr{B}} - E_\txr{in} - \epsilon^{(0)}_{pn})
\Psi^{(-)}_{p n \txr{B}, \vK_2} (\vb*{r}, \vR_2) = 0,
\end{align}
in which $E_\txr{in}$ is the incident energy of $d$ in the $d$-B c.m.~system and $\epsilon^{(0)}_{pn}$ is the eigenenergy of deuteron. The three-body Hamiltonian $\hat{H}_{p n \txr{B}}$ is defined by
\begin{align}
\hat{H}_{p n \txr{B}} \equiv
\hat{T}_{\vR_2}
+ U_{p \txr{B}} (R_{2+})
+ U_{n \txr{B}} (R_{2-})
+ \hat{h}_{pn},
\label{eq5}
\end{align}
in which $\hat{T}_{\vR_2}$ is the kinetic-energy operator with respect to $\vR_2$, $U_{p \txr{B}}$ ($U_{n \txr{B}}$) is the distorting potential for the proton (neutron) by B with
\begin{align}
R_{2 \pm} \equiv \abs{\vR_2 \pm \frac{1}{2} \vb*{r}},
\end{align}
and $\hat{h}_{pn}$ is the internal Hamiltonian of the $pn$ system. In CDCC, we expand $\Psi_{p n \txr{B}, \vK_2}$ as
\begin{align}
\Psi^{(-)}_{p n \txr{B}, \vK_2} (\vb*{r}, \vR_2) =
\sum^{c_\txr{max}}_{c = 0}
\chi^{(c) (-)}_{pn, \vK_2} (\vR_2)
\Phi^{(c)}_{pn} (\vb*{r}),
\label{eq7}
\end{align}
in which $\Phi^{(c)}_{pn}$ is the internal wave function of the deuteron for $c = 0$ ($\Phi^{(0)}_{pn} = \Phi_d$) or the $pn$ system in discretized-continuum states for $c \ne 0$. For each $c$, $\Phi^{(c)}_{pn}$ satisfies the eigenvalue equation
\begin{align}
\hat{h}_{pn} \Phi^{(c)}_{pn} (\vb*{r}) =
\epsilon^{(c)}_{pn} \Phi^{(c)}_{pn} (\vb*{r})
\end{align}
with $\epsilon^{(c)}_{pn}$ being the eigenenergy of the $pn$ system in the $c$th state. Equation~\eqref{eq7} means that the three-body scattering wave is expanded in terms of the set of the eigenstates $\Phi^{(c)}_{pn}$ of $\hat{h}_{pn}$, which is assumed to form a complete set in the model space relevant to the physics observables of interests. The expansion ``coefficients'' $\chi^{(c)}_{pn, \vK_2}$ physically represent the scattering waves between the c.m.~of the $pn$ system in the $c$th states and B. The boundary condition on the wave function~\eqref{eq7} is given as
\begin{widetext}
\begin{align}
\Psi^{(-)}_{p n \txr{B}, \vK_2} (\vb*{r}, \vR_2) \rightarrow
\frac{1}{(2 \pi)^{3/2}} \qty[\exp(i \vK_{2, c} \cdot \vR_2)
\Phi^{(c)}_{pn} (\vb*{r}) \delta_{c 0} +
\sum^{c_\txr{max}}_{c = 0} f^{(c)*}_{p n \txr{B}} \qty(\hat{\vK}_{2, c})
\frac{\exp(- i K_{2, c} R_2)}{R_2} \Phi^{(c)}_{pn} (\vb*{r})] \quad
(R_2 \rightarrow \infty).
\end{align}
\end{widetext}
Here, $\vK_{2, c}$ is the asymptotic momentum of the c.m.~of the $pn$ system in channel $c$ and $f^{(c)}_{p n \txr{B}}$ is the corresponding scattering amplitude; $\vK_{2, 0}$ is denoted by $\vK_2$ for simplicity. The magnitudes of $\vK_{2, c}$ are determined with the energy conservation by
\begin{align}
\frac{\hbar^2}{2 \mu_{2 \txr{B}}} K^2_{2, c} +
\epsilon^{(c)}_{pn} =
\frac{\hbar^2}{2 \mu_{2 \txr{B}}} K^2_2 +
\epsilon^{(0)}_{pn}.
\label{eq10}
\end{align}

Inserting Eq.~\eqref{eq7} into Eq.~\eqref{eq1}, the transition matrix becomes
\begin{align}
T =
&\sum^{c_\txr{max}}_{c = 0}
\bra{\chi^{(-)}_{1, \vK_1} (\vR_1)
\chi^{(c) (-)}_{pn, \vK_2} (\vR_2)
\Phi^{(c)}_{pn} (\vb*{r})}
[t_{1p} (\vb*{s}_{1p}) \nn
&\quad + t_{1n} (\vb*{s}_{1n})]
\hat{\mathcal{A}}_\txr{elm}
\ket{\chi^{(+)}_{0, \vK_0} (\vR_0)
\varphi_d (\vR_2) \Phi_d (\vb*{r})}.
\label{eq11}
\end{align}
To express all coordinates relevant to the transition matrix~\eqref{eq11} by $\vR$ (the c.m.~coordinate of the 1-2 system for B) and $\vb*{s}$, the asymptotic momentum approximation (AMA) \cite{KazukiYoshida16, TWakasa17} to the distorted waves is adopted. In this approximation, the propagation of each distorted wave from a point $\vb*{x}$ to another point $\vb*{x} + \Delta \vb*{x}$ is expressed by the plane wave with the asymptotic momentum, i.e.,
\begin{align}
\chi_{\vK} (\vb*{x} + \Delta \vb*{x}) \approx
\chi_{\vK} (\vb*{x}) \txr{e}^{i \vK \cdot \Delta \vb*{x}},
\end{align}
with the assumption that $\Delta \vb*{x}$ is small. This approximation has been standardly adopted in describing ($p,2p$) reactions \cite{NSChant77, NSChant83, YKudo86, YKudo88, NKanayama90, YKudo93, KOgata15, TWakasa17}, and the reaction calculations employing the AMA have been successfully used as a nuclear spectroscopic tool. The coordinates $\vR_0$, $\vR_1$, and $\vR_2$ can be expressed in terms of $\vR$ and $\vb*{s}$ as
\begin{align}
\vR_0 = \vR - \alpha_R \vR + \alpha_{0s} \vb*{s},
\end{align}
\begin{align}
\vR_1 = \vR + \alpha_{1s} \vb*{s}, \quad
\vR_2 = \vR - \alpha_{2s} \vb*{s},
\end{align}
in which
\begin{align}
\alpha_R = \frac{2}{A}, \quad
\alpha_{0s} = \frac{2}{3} \frac{A + 1}{A}, \quad
\alpha_{1s} = \frac{2}{3}, \quad
\alpha_{2s} = \frac{1}{3}.
\end{align}
Equation~\eqref{eq11} is then reduced to
\begin{align}
T =
\sum^{c_\txr{max}}_{c = 0}
\tilde{t}^{(c)}_{12} (\vkap'_c, \vkap)
\tilde{T}^{(c)},
\label{eq16}
\end{align}
in which
\begin{widetext}
\begin{align}
\tilde{t}^{(c)}_{12} (\vkap'_c, \vkap) \equiv
\mel**{\txr{e}^{i \vkap'_c \cdot \vb*{s}}
\Phi^{(c)}_{pn} (\vb*{r})
}{[t_{1p} (\vb*{s}_{1p}) + t_{1n} (\vb*{s}_{1n})]
\hat{\mathcal{A}}_\txr{elm}
}{\txr{e}^{i \vkap \cdot \vb*{s}}
\Phi_d (\vb*{r})},
\label{eq17}
\end{align}
\begin{align}
\tilde{T}^{(c)} \equiv
\int \dd{\vR}
\chi^{(-) *}_{1, \vK_1} (\vR)
\chi^{(c) (-) *}_{2, \vK_2} (\vR)
\chi^{(+)}_{0, \vK_0} (\vR)
\txr{e}^{- i \alpha_R \vK_0 \cdot \vR}
\varphi_d (\vR).
\end{align}
\end{widetext}
Here, $\tilde{t}^{(c)}_{12}$ is the transition matrix for the $p$-$d$ elastic scattering for $c = 0$ and $d(p,p)pn$ reaction for $c \ne 0$. It means that the reaction process is described with an impulse picture employing a nucleon-nucleon effective interaction. The explicit form of Eq.~\eqref{eq17} considering the $S$-waves as the bound and breakup states of deuteron is given in Appendix C of Ref.~\cite{YChazono22}. $\vkap$ ($\vkap'_c$) is the relative momentum between the proton and the $pn$ system in the initial (final) state defined by
\begin{align}
\vkap \equiv \alpha_{0s} \vK_0 - \alpha_{2s} \vK_d, \quad
\vkap'_c \equiv \alpha_{1s} \vK_1 - \alpha_{2s} \vK_{2, c},
\end{align}
with $\vK_d$ being the momentum of the Fermi motion of the deuteron inside A. Following Ref.~\cite{TWakasa17}, we assume that the momentum conservation of the two colliding particles in the elementary process holds, i.e.,
\begin{align}
\frac{A + 1}{A} \vK_0 + \vK_d \approx
\vK_1 + \vK_2.
\end{align}
$\vkap$ is then determined by
\begin{align}
\vkap =
\frac{A + 1}{A} \vK_0 - \alpha_{2s} (\vK_1 + \vK_2)
\end{align}
and $\tilde{t}^{(c)}_{12}$ becomes independent of $\vK_d$. We make the on-the-energy-shell (on-shell) approximation of the final-state prescription, i.e., $\vkap = \kappa'_0 \hat{\vkap}$, the transition matrix~\eqref{eq17} is related to the differential cross section of the $p$-$d$ elastic scattering (for $c = 0$) and $d(p,p)pn$ reaction (for $c \ne 0$) as
\begin{align}
\frac{\dd{\sigma^{(c)}_{pd}}}{\dd{\Omega_{pd}}} (\theta_{pd}, T_{pd}) =
C_\txr{corr}
\qty(\frac{\mu_{pd}}{2 \pi \hbar^2})^2 \frac{1}{6}
\abs{\tilde{t}^{(c)}_{12} (\vkap'_c, \vkap)}^2.
\label{eq22}
\end{align}
Here, $\mu_{pd}$ is the reduced energy of the $p$-$d$ system, and $\theta_{pd}$ and $T_{pd}$ are the scattering angle and kinetic energy, respectively. It should be noted that because nucleon-nucleon interactions are parametrized to reproduce the nucleon-nucleon scattering data, there is no guarantee for the interactions to reproduce the experimental data of the reactions of the $p$-$d$ system. Thus, we have introduced the correction factor $C_\txr{corr}$, which is determined to reproduce the $p$-$d$ elastic scattering data at forward angles. The same $C_\txr{corr}$ is used in the calculation of the $d(p,p)pn$ reaction.

Using Eq.~\eqref{eq16}, the triple differential cross section (TDX) of A($p,pd$)B is obtained as
\begin{widetext}
\begin{align}
\frac{\txr{d}^3 \sigma^\trL}
{\txr{d} E^\trL_1 \txr{d} \Omega^\trL_1 \txr{d} \Omega^\trL_2} =
\frac{(2 \pi)^4}{\hbar v_\txr{i}}
\frac{E_1 E_2 E_\txr{B}}{E^\trL_1 E^\trL_2 E^\trL_\txr{B}}
F^\trL_\txr{kin}
\frac{1}{2 s_0 + 1} \frac{1}{2 s_d + 1}
\abs{\sum^{c_\txr{max}}_{c = 0}
\tilde{t}^{(c)}_{12} (\vkap'_c, \vkap)
\tilde{T}^{(c)}}^2,
\label{eq23}
\end{align}
\end{widetext}
which is the cross section specifying the total energy $E^\trL_1$ and emission direction $\Omega^\trL_1$ of particle 1 and the direction $\Omega^\trL_2$ of particle 2. $v_\txr{i}$ is the magnitude of the relative velocity between particle 0 and A of Lorentz invariance, which is given by
\begin{align}
v_\txr{i} \equiv c^2 \frac{\hbar K^\trL_0}{E^\trL_0},
\end{align}
and $E_j$ ($j = 0, 1, 2$, or B) is the total energy of particle or nucleus $j$. $F^\trL_\txr{kin}$ is the phase volume given by
\begin{align}
F^\trL_\txr{kin} \equiv
\frac{E^\trL_1 K^\trL_1 E^\trL_2 K^\trL_2}{(\hbar c)^4}
\qty[1 + \frac{E^\trL_2}{E^\trL_\txr{B}}
+ \frac{E^\trL_2}{E^\trL_\txr{B}}
\frac{\vK^\trL_\txr{X} \cdot \vK^\trL_2}{(K^\trL_2)^2}]^{-1}
\end{align}
with $\vK^\trL_\txr{X} \equiv \vK^\trL_1 - \vK^\trL_0 - \vK^\trL_\txr{B}$. $s_0 = 1/2$ and $s_d = 1$ are the spins of particle 0 and the deuteron, respectively.

When only the elastic channel ($c=0$) is considered for particle 2, Eq.~\eqref{eq7} is reduced to
\begin{align}
\Psi^{(-)}_{p n \txr{B}, \vK_2} (\vb*{r}, \vR_2) \rightarrow
\chi^{(0) (-)}_{2, \vK_2} (\vR_2)
\Phi_d (\vb*{r}).
\label{eq26}
\end{align}
In this case, only the $p$-$d$ elastic scattering is involved as the elementary process of the ($p,pd$) reaction and the TDX~\eqref{eq23} becomes
\begin{widetext}
\begin{align}
\frac{\txr{d}^3 \sigma^{(\txr{NB}) \trL}}
{\txr{d} E^\trL_1 \txr{d} \Omega^\trL_1 \txr{d} \Omega^\trL_2} =
\frac{(2 \pi)^4}{\hbar v_\txr{i}}
\frac{E_1 E_2 E_\txr{B}}{E^\trL_1 E^\trL_2 E^\trL_\txr{B}}
F^\trL_\txr{kin}
\qty(\frac{2 \pi \hbar^2}{\mu_{pd}})^2
\frac{\dd{\sigma^{(0)}_{pd}}}{\dd{\Omega_{pd}}} (\theta_{pd}, T_{pd})
\abs{\tilde{T}^{(\txr{NB})}}^2.
\label{eq27}
\end{align}
\end{widetext}
The superscript (NB), the abbreviation of ``no breakup,'' is put to emphasize that the TDX~\eqref{eq27} does not include breakup of deuteron at all. We call this approach NB-CDCCIA below.

\section{Result and discussion \label{sec3}}

\subsection{Numerical input for the elementary process \label{subsec3A}}
Before discussing the results of CDCCIA, it should be tested how well Eq.~\eqref{eq17} describes the elementary processes of the ($p,pd$) reaction, i.e., the $p$-$d$ elastic scattering and $d(p,p)pn$ reaction. The cross section~\eqref{eq22} is calculated at the incident energies $T^\trL_p = 120$, $135$, $155$, $170$, $190$, and $250$~MeV.

The Franey-Love parameter set \cite{MAFraney85} is employed for the nucleon-nucleon effective interactions in Eq.~\eqref{eq17}. The Coulomb interaction between the two protons is ignored because its effect is known to be restricted at very forward angles. In the present calculation, we include only the $S$-waves in the bound and breakup states of the deuteron. The reason is that, as described in Appendix C of Ref.~\cite{YChazono22}, the actual computation of Eq.~\eqref{eq17} is rather difficult even when only the $S$-waves are considered, and that the extension to the $P$- and $D$-waves involves additional difficulties. The breakup state is discretized with the pseudostate method \cite{TMatsumoto03}, in which the wave functions of the discretized-continuum states are obtained by diagonalizing the internal Hamiltonian $\hat{h}_{pn}$ of the $pn$ system. The potential between the proton and neutron in $\hat{h}_{pn}$ is assumed to have a one-range Gaussian form
\begin{align}
v_{pn} (r) = v_0 \exp[- \qty(\frac{r}{r_0})^2],
\label{eq28}
\end{align}
in which $v_0 = -72.15$~MeV and $r_0 = 1.484$~fm \cite{TOhmura70} are used to reproduce the binding energy and radius of the deuteron. In the calculation of the $d(p,p)pn$ reaction, the cross sections~\eqref{eq22} are summed up to the $pn$ eigenenergies of $25$~MeV. The correction factor $C_\txr{corr}$ is determined to reproduce the elastic scattering data between $15^\circ$ and $80^\circ$ at each incident energy.

\subsection{$p$-$d$ elastic scattering and breakup reaction \label{subsec3B}}

\begin{figure*}[htbp]
\begin{center}
\includegraphics[width=0.42\textwidth]{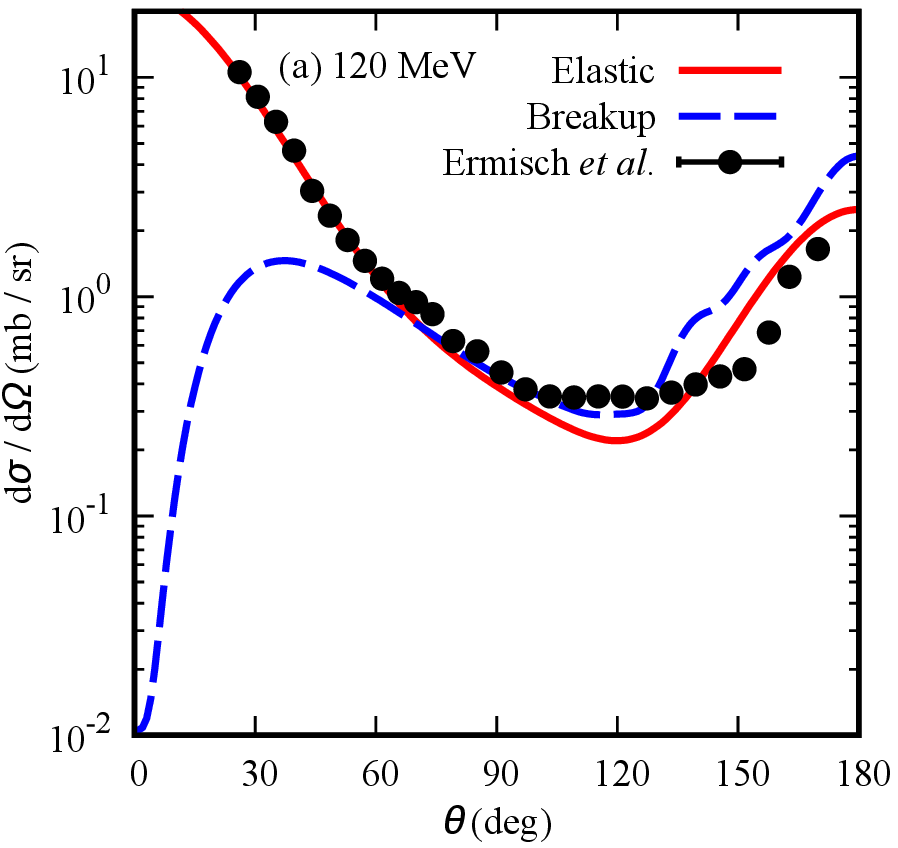}
\hspace{1.5cm}
\includegraphics[width=0.42\textwidth]{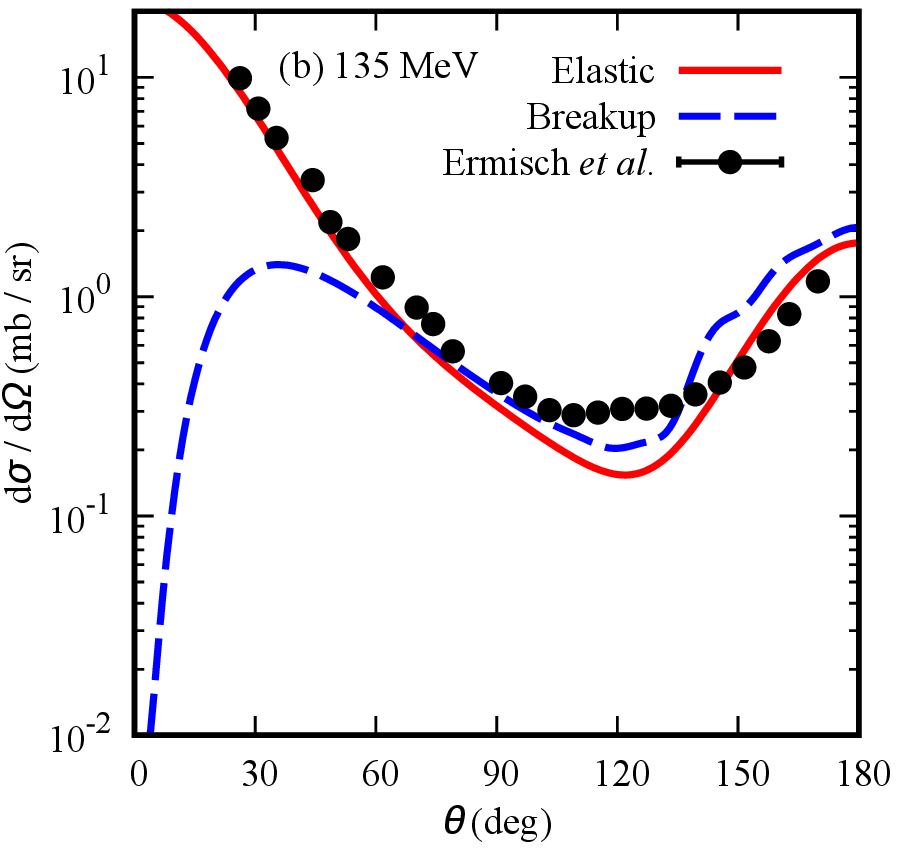}
\vspace{0.5cm}\\
\includegraphics[width=0.42\textwidth]{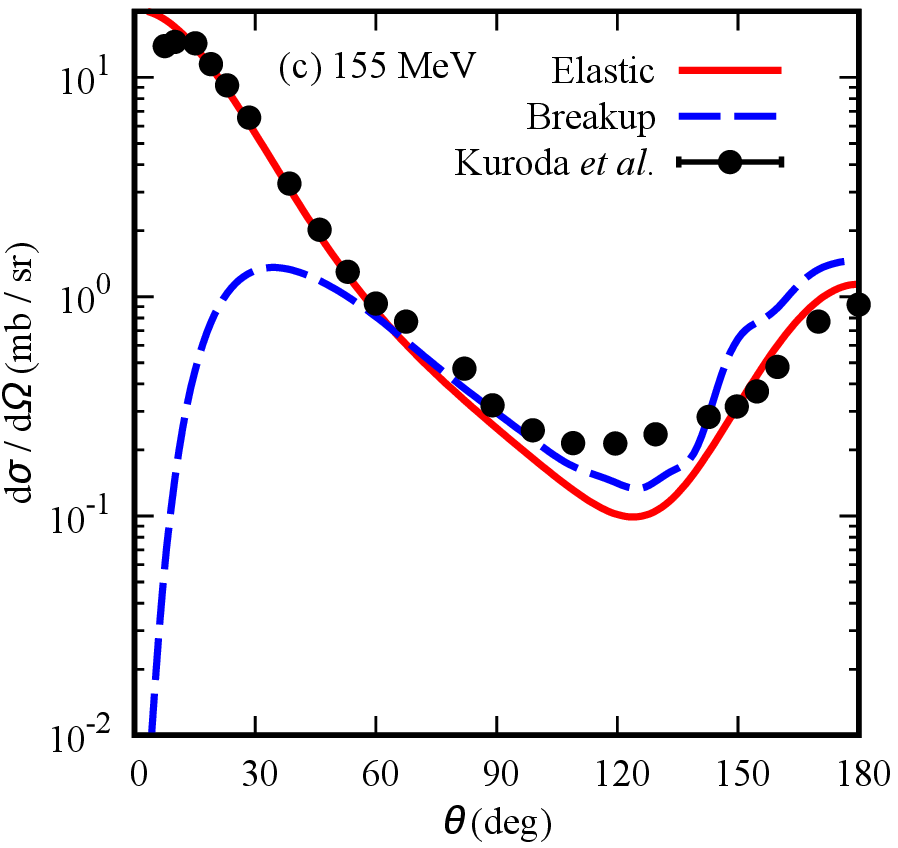}
\hspace{1.5cm}
\includegraphics[width=0.42\textwidth]{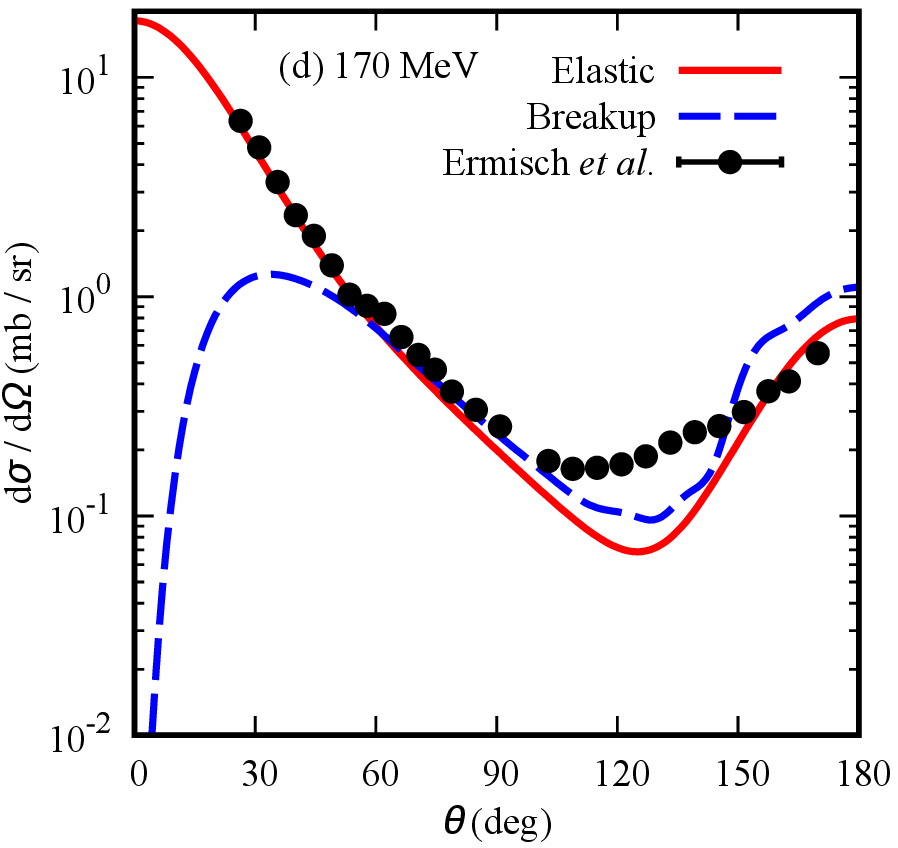}
\vspace{0.5cm}\\
\includegraphics[width=0.42\textwidth]{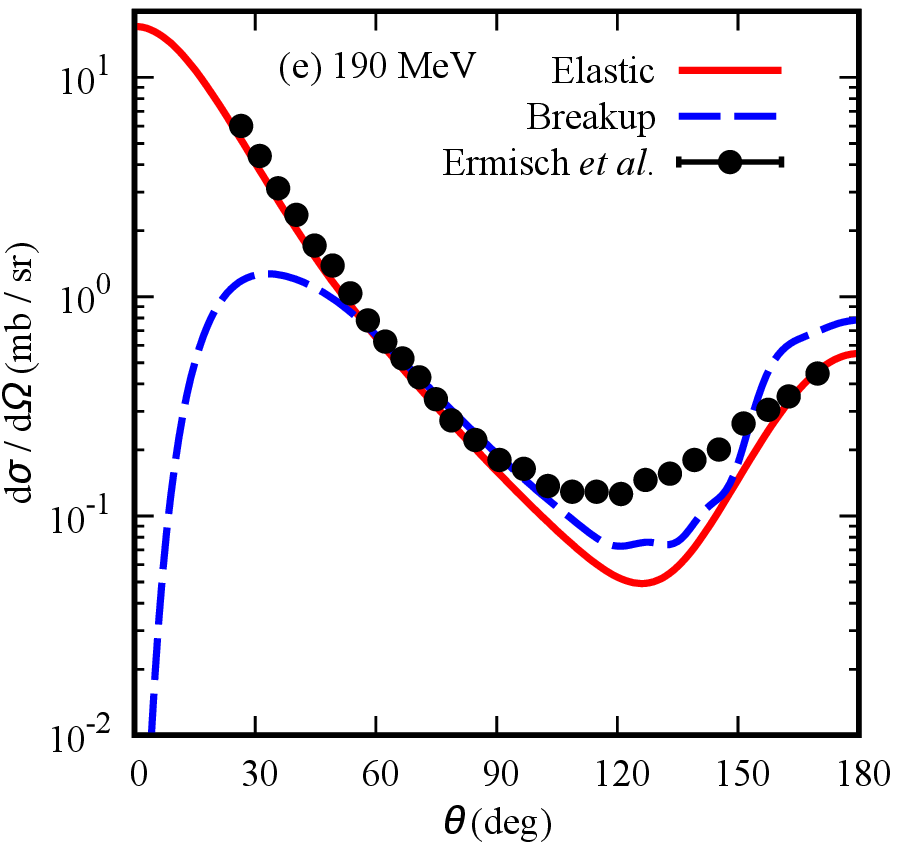}
\hspace{1.5cm}
\includegraphics[width=0.42\textwidth]{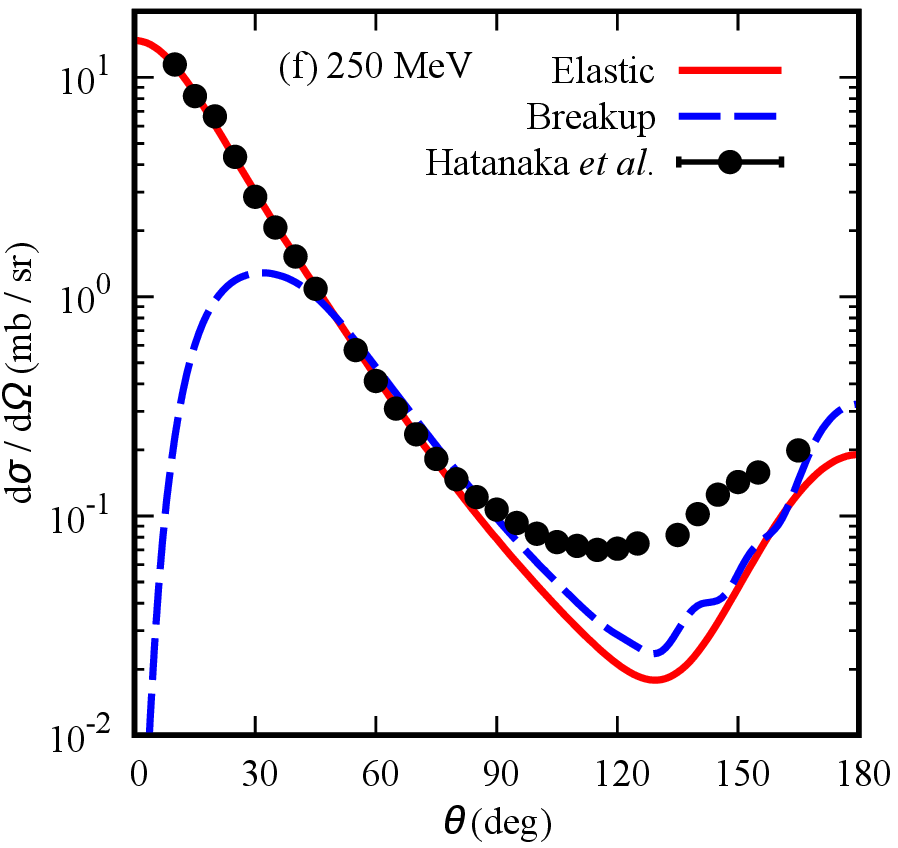}
\caption{Differential cross sections of the $p$-$d$ elastic scattering and $d(p,p)pn$ reaction at $T^\txr{L}_p =$ (a)~$120$, (b)~$135$, (c)~$155$, (d)~$170$, (e)~$190$, and (f)~$250$~MeV. The horizontal axis is the scattering angle in the c.m.~frame of the $p$-$d$ system. The solid (dashed) line shows the numerical result of elastic scattering (breakup reaction) in each panel. The dots are the experimental data of the $p$-$d$ elastic scattering taken from Ref.~\cite{KErmisch05} ($120$, $135$, $170$, and $190$~MeV), Ref.~\cite{KKuroda64} ($155$~MeV), and Ref.~\cite{KHatanaka02} ($250$~MeV). \label{fig2}}
\end{center}
\end{figure*}

Figure~\ref{fig2} shows the differential cross sections of the $p$-$d$ elastic scattering and $d(p,p)pn$ reaction as a function of the scattering angle in the c.m.~frame of the $p$-$d$ system. Panels (a), (b), (c), (d), (e), and (f) correspond to $120$, $135$, $155$, $170$, $190$, and $250$~MeV, respectively. In each panel, the solid and dashed lines represent the numerical results of the elastic scattering and $d(p,p)pn$ reaction, respectively. The experimental data of the $p$-$d$ elastic scattering denoted by the dots are taken from Ref.~\cite{KErmisch05} ($120$, $135$, $170$, and $190$~MeV), Ref.~\cite{KKuroda64} ($155$~MeV), and Ref.~\cite{KHatanaka02} ($250$~MeV).

\begin{table}[htbp]
\begin{center}
\caption{Correction factor $C_\txr{corr}$ \label{tab1}}
\begin{tabular}{ccccccc}
\hline\hline
$T^\txr{L}_p$~(MeV) & $120$  & $135$  & $155$  & $170$  & $190$  & $250$  \\
\hline
$C_\txr{corr}$      & $0.70$ & $0.70$ & $0.71$ & $0.68$ & $0.69$ & $0.68$ \\
\hline\hline
\end{tabular}
\end{center}
\end{table}

One sees that the calculated $p$-$d$ elastic scattering cross sections (solid lines) reproduce the forward-angle ($\lesssim 90^\circ$) data at all energies considered. The obtained $C_\txr{corr}$ are shown in Table~\ref{tab1}; it is found that the energy dependence of $C_\txr{corr}$ is weak enough to be regarded as a constant between 120 and 250 MeV. Thus, the average value of $0.69$ is used in the ($p,pd$) calculation. The experimental data at backward angles ($\gtrsim 150^\circ$) are also described well by the calculated results, though some undershooting is found at $250$~MeV. In the backward scattering, it is known that the pickup process plays a dominant role, in which the incident proton picks up the neutron in the deuteron. As the incident energy increases, the neutron inside the deuteron needs to have a high momentum to be picked up by the high energy proton, as indicated by the so-called momentum matching condition. The $pn$ potential~\eqref{eq28} cannot generate such a high-momentum component, which is expected to be the reason for the undershooting of the data at backward angles at $250$~MeV. Further improvement of this approach will be needed to be applied to the study of the ($p,pd$) reaction with the backward emission of the proton \cite{STerashima18}; the use of a realistic $pn$ interaction will be necessary. However, it is beyond the scope of this article because the kinematics employed in Ref.~\cite{STerashima18} are completely different from those discussed in this study, in which the proton is emitted to forward angles.

From about $90^\circ$ to $150^\circ$, one finds that the calculations underestimate the data. It is well known that three-body interactions, which are not included in the present calculation, are essential to reproduce the observables of the nucleon-deuteron elastic scattering at those angles \cite{HWitala98, SNemoto98, WPAbfalterer98, HWitala99, KSekiguchi02, KErmisch05}. It should be noted that the $p$-$d$ scattering at forward angles between $65^\circ$ and $80^\circ$ mainly contributes to the ($p,pd$) process with the kinematics condition adopted in this study. Thus, the present model can be employed without considering such an issue.

Unfortunately, there are no experimental data of the $d(p,p)pn$ reaction that can directly be compared with the calculated results (dashed lines) shown in Fig.~\ref{fig2}. Although it is difficult to draw a clear conclusion, one may expect that the present description of the $d(p,p)pn$ process is, at least to some extent, reliable. This is because the elastic process is described well at forward angles as seen above and the nucleon-nucleon effective interaction is independent of whether the $pn$ system is in the bound state (deuteron) or continuum state. Thus, Eq.~\eqref{eq17} is employed to evaluate the contribution of the deuteron breakup in the elementary process to the TDX of the ($p,pd$) reaction.

\subsection{Numerical input for A($p,pd$)B reaction \label{subsec3C}}
We consider three A($p,pd$)B reactions at $250$~MeV in normal kinematics. The target (residual) nuclei A (B) and their states are listed in Table~\ref{tab2}. The emission angles of the proton (deuteron) in the final state are fixed at $47.1^\circ~(48.8^\circ)$ for $^{16}$O, $49.0^\circ~(48.6^\circ)$ for $^{40}$Ca, and $48.9^\circ~(48.6^\circ)$ for $^{56}$Ni. As mentioned above, the asymptotic momenta of particles 0, 1, and 2 are kept on the $z$-$x$ plane. The azimuthal angles of the proton and deuteron are set to $0^\circ$ and $180^\circ$, respectively.

\begin{table}[htbp]
\begin{center}
\caption{Target, residue, and deuteron orbit \label{tab2}}
\begin{tabular}{ccccccc}
\hline\hline
A         & State   & $\quad$ & B         & State   & $\quad$ & Orbit \\
\hline
$^{16}$O  & $0^+_1$ & $\quad$ & $^{14}$N  & $1^+_2$ & $\quad$ & $1S$  \\
$^{40}$Ca & $0^+_1$ & $\quad$ & $^{38}$K  & $1^+_1$ & $\quad$ & $2S$  \\
$^{56}$Ni & $0^+_1$ & $\quad$ & $^{54}$Co & $1^+_1$ & $\quad$ & $3S$  \\
\hline\hline
\end{tabular}
\end{center}
\end{table}

In the initial state, the deuteron is assumed to be bound in the $nS$ orbit with the number of nodes $n$ (see Table.~\ref{tab2}). The binding potential is assumed to be a central Woods-Saxon potential
\begin{align}
V (R) = - V_0 \frac{1}{1 + \exp[(R - R_0) / a_0]}
\end{align}
with the radial parameter $R_0 = 1.41 \times B^{1/3}$~fm and the diffuseness parameter $a_0 = 0.65$~fm; these are taken from Ref.~\cite{CSamanta82}. The depth $V_0$ of the potential is determined to reproduce the deuteron separation energies of A of $24.68$ ($^{16}$O), $19.60$ ($^{40}$Ca), and $19.97$ ($^{56}$Ni) MeV. The Coulomb part is constructed by assuming a uniformly charged sphere with the radius $R_\txr{C} = 1.41 \times B^{1/3}$~fm.

The EDAD1 parameter set of the Dirac phenomenology \cite{SHama90, EDCooper93, EDCooper09} is adopted for the distorting potentials of the $p$-A, $p$-B, and $n$-B systems. The Coulomb potential in each potential is constructed in the same way as in the previous paragraph with $R_\txr{C} = 1.2 \times C^{1/3}$~fm ($C = A~\txr{or}~B$). The nonlocality correction to the distorted waves of the incoming and outgoing protons are taken into account by including the Darwin factor \cite{SHama90, LGArnold81}, while that for deuteron is not considered because it is difficult to treat it in a consistent manner with the calculation of $\chi_{0, \vK_0}$ and $\chi_{1, \vK_1}$ \cite{NKTimofeyuk13a, NKTimofeyuk13b, RCJohnson14}.

In the CDCC calculation, the breakup state of the deuteron is discretized with the pseudostate method \cite{TMatsumoto03} as in Sec.~\ref{subsec3A}. The $pn$ continuum states of $\ell = 0, 2,~\txr{and}~4$ are included with $k_\txr{max} = 2.1$~fm$^{-1}$, in which $\ell$ is the orbital angular momentum between $p$ and $n$, and $k_\txr{max}$ is the maximum $p$-$n$ linear momentum. The model space is determined so that the calculated cross section of the $d$-B elastic scattering converged. In the calculation of Eq.~\eqref{eq23}, as the first step of the CDCCIA study on the $(p,pd)$ reaction, we include only the ground state and the $S$-wave discretized-continuum states up to their eigenenergies of $25$~MeV. It should be noted that in this case, the effect of Coulomb breakup, which is ignored in the preset study, is expected to be negligible.

\subsection{TDX of A($p,pd$)B reaction \label{subsec3D}}

\begin{figure*}[htbp]
\begin{center}
\includegraphics[width=0.84\textwidth]{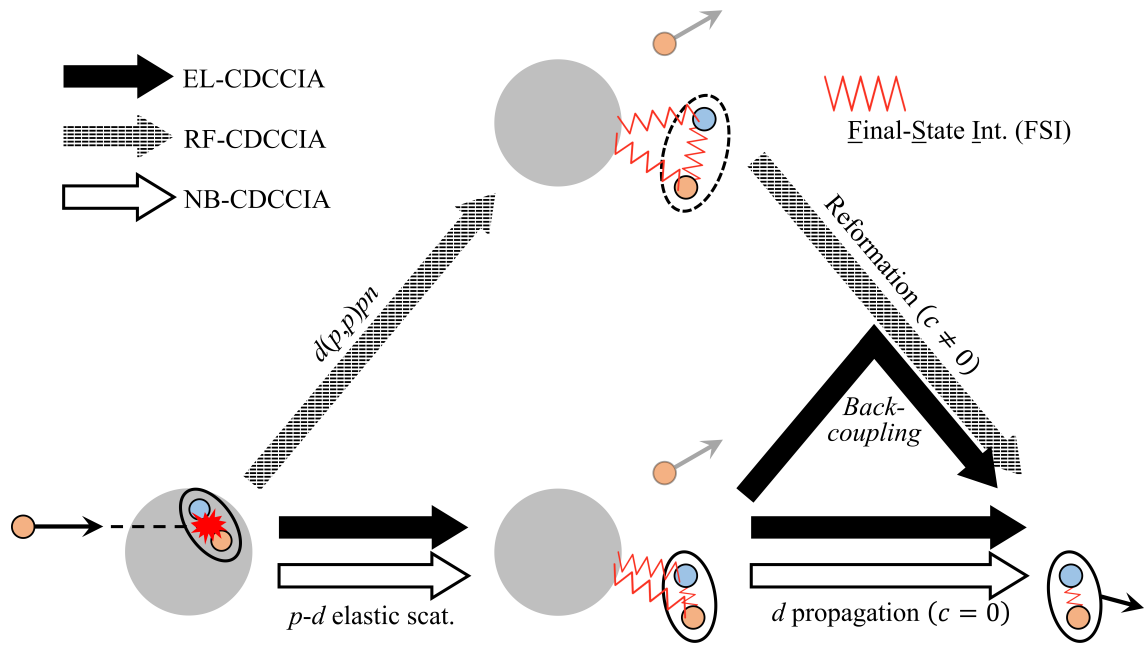}
\caption{Schematic illustration of the reaction processes included in the present CDCCIA calculation. EL- and RF-CDCCIA are CDCCIA considering the elastic and breakup channels of the wave function~\eqref{eq7}, respectively. \label{fig3}}
\end{center}
\end{figure*}

In Fig.~\ref{fig3}, we show the schematic illustration of the reaction processes included in the present CDCCIA calculation. The left half shows the branches regarding the elementary process of the reaction, which are described with the approach discussed in Sec.~\ref{subsec3B}. The right half shows the role of the FSIs discussed below. The paths represented by the filled and meshed arrows show CDCCIA taking the $c=0$ and $c \ne 0$ component(s) in Eq.~\eqref{eq23}, respectively. The former and latter are referred to as EL- and RF-CDCCIA; EL and RF are, respectively, the abbreviations of ``elastic'' and ``reformation.'' The path corresponding to NB-CDCCIA~\eqref{eq27} is denoted by the open arrows.

\begin{figure}[htbp]
\begin{center}
\includegraphics[width=0.42\textwidth]{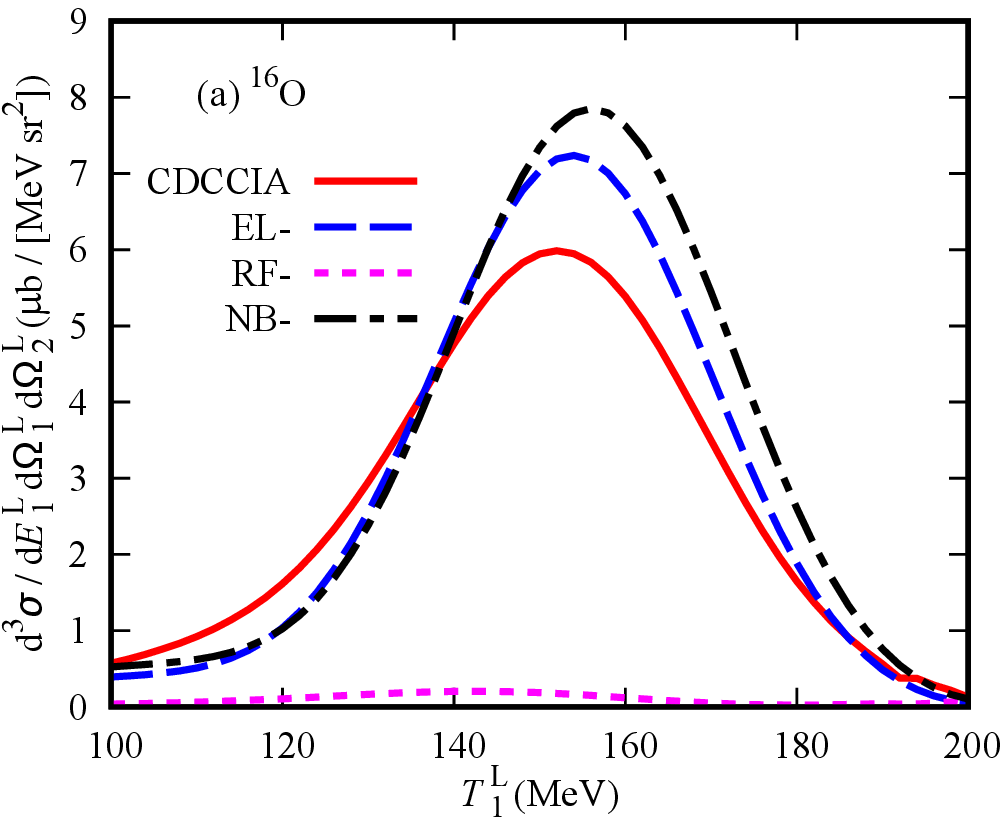}
\vspace{0.2cm}\\
\includegraphics[width=0.42\textwidth]{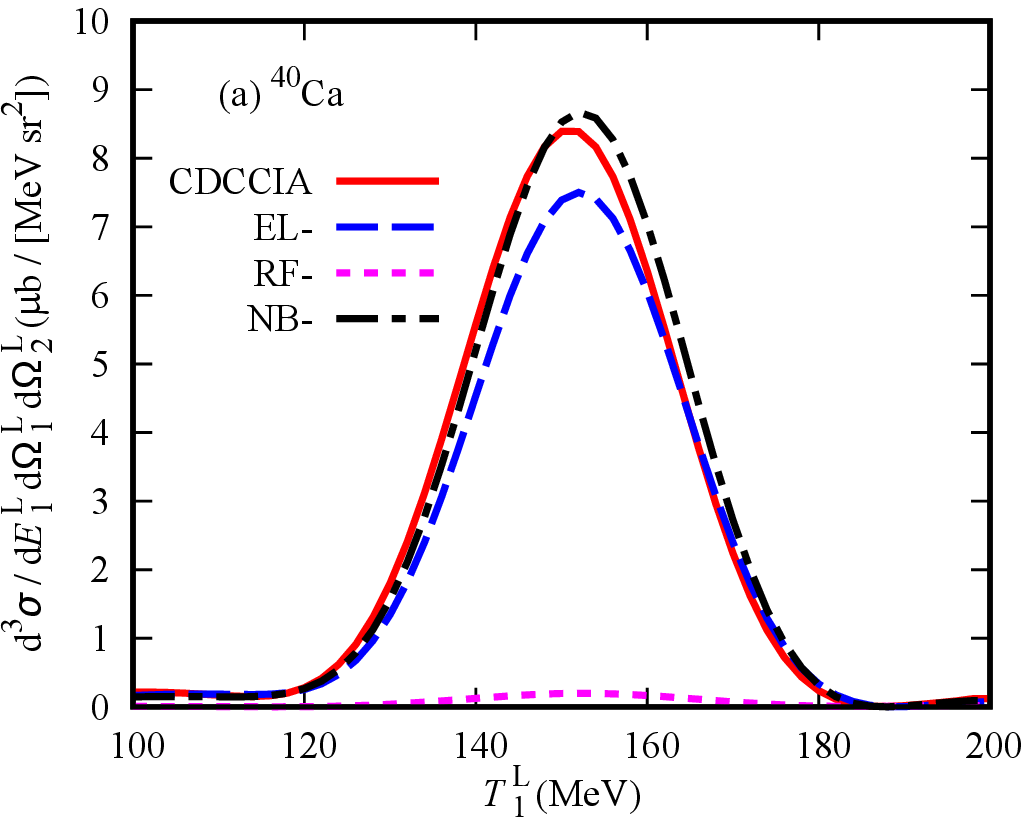}
\vspace{0.2cm}\\
\includegraphics[width=0.42\textwidth]{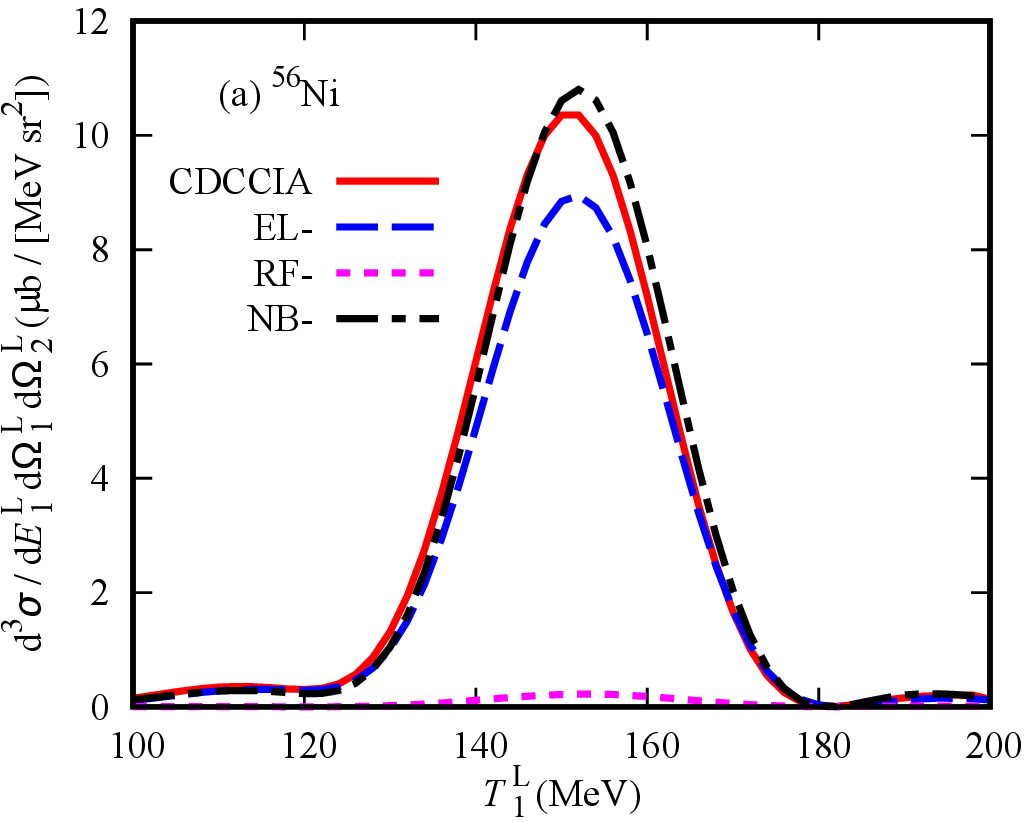}
\caption{Triple differential cross sections (TDXs) of the ($p,pd$) reactions at $250$~MeV. The target nuclei are (a) $^{16}$O, (b) $^{40}$Ca, and (c) $^{56}$Ni. The solid lines show the CDCCIA results of Eq.~\eqref{eq23}. The dashed (dotted) lines show the TDXs obtained by taking the $c=0$  ($c \ne 0$) component(s) in Eq.~\eqref{eq23}, referred to as EL-CDCCIA (RF-CDCCIA). The dot-dashed lines are the results without any deuteron breakup, i.e., NB-CDCCIA of Eq.~\eqref{eq27}. \label{fig4}}
\end{center}
\end{figure}

The TDX of the $^{16}$O($p,pd$)$^{14}$N at $250$~MeV are shown in Fig.~\ref{fig4}(a) as a function of the kinetic energy $T^\trL_1$ of the outgoing proton (particle 1). $T^\trL_1 \approx 150$~MeV satisfies the recoilless condition, in which the momentum $\vK_\txr{B}$ of B is almost zero. The CDCCIA result of Eq.~\eqref{eq23} is represented by the solid line. The dashed and dotted lines show the TDXs calculated with EL- and RF-CDCCIA, respectively. The result of NB-CDCCIA is also shown by the dot-dashed line. Figures~\ref{fig4}(b) and 4(c) are the same as Fig.~\ref{fig4}(a) but for (b) $^{40}$Ca($p,pd$)$^{38}$K and (c) $^{56}$Co($p,pd$)$^{54}$Ni reactions.

First, we compare the TDXs of CDCCIA with those of EL-CDCCIA. The difference between the two is whether the deuteron breakup channels ($c \ne 0$) are included in the calculation or not. Since these channels affect the TDX as the deuteron reformation process by the FSIs (see Fig.~\ref{fig3}), the difference between the two TDXs implies a role of this process. In the $^{16}$O($p,pd$)$^{14}$N reaction (Fig.~\ref{fig4}(a)), one can see that the deuteron reformation process decreases the TDX by about $21$~\% around the peak. In contrast, this process increases the TDXs of the $^{40}$Ca($p,pd$)$^{38}$K (Fig.~\ref{fig4}(b)) and $^{56}$Ni($p,pd$)$^{54}$Co (Fig.~\ref{fig4}(c)) by about $11$ and $14$~\%, respectively. The latter two results are consistent with the intuition that the cross section will increase because the number of channels in which the $pn$ system forms deuteron in the end increases. On the other hand, the decrease in the TDX from EL-CDCCIA (dashed line) to CDCCIA (solid line) in Fig.~\ref{fig4}(a) seems to contradict this intuition. This phenomenon is caused by the interference between the elastic and breakup channels of deuteron. As shown in Eq.~\eqref{eq23}, the matrix elements of the elastic and breakup channels are summed up coherently in the CDCCIA calculation. Therefore, the calculated TDX includes the interference between those channels. Since the interference does not always make a constructive contribution, the TDX of CDCCIA can be smaller than that of EL-CDCCIA due to the deuteron reformation. From Fig.~\ref{fig4}, it can be also seen that the TDXs of RF-CDCCIA (dotted lines), the cross sections of the deuteron reformation process in the ($p,pd$) reaction, is almost negligible. The difference in the TDX between CDCCIA and EL-CDCCIA is, however, larger than the TDX height of RF-CDCCIA. Thus, it indicates that the deuteron reformation significantly changes the TDX of the ($p,pd$) reaction through the interference. Therefore, it is important to include this process in the calculation of ($p,pd$) cross sections when the results are compared with experimental data.

Next, we compare the TDXs of EL-CDCCIA (dashed lines) and those of NB-CDCCIA (dot-dashed lines). The former correspond to the process in which the deuteron in A is knocked out as it is, but its breakup due to the FSI is taken into account by using CDCC. In other words, the knocked-out deuteron can be broken up during the scattering process with B, but it backs to the ground state (deuteron) in the end; this is called the \textit{back-coupling} effect (see also Fig.~\ref{fig3}). On the other hand, the latter are obtained by the calculation in which no breakup state is considered, i.e., the back-coupling effect is also absent. In Fig.~\ref{fig4}, one finds that the TDXs of EL-CDCCIA are smaller than those of NB-CDCCIA in the almost whole region of $T^\trL_1$. These differences indicate the importance of the back-coupling effect. A qualitative explanation of the role of the back-coupling effect is given in terms of the flux. In the EL-CDCCIA calculation, through the CDCC calculation, the flux of the elastic (incident) channel is distributed to the breakup ones and reduced because of the conservation of flux. This flux-loss is obviously absent in the NB-CDCCIA calculation. As a result, the amplitude of $\chi^{(0)}_{2, \vK_2}$ used in EL-CDCCIA becomes smaller than that in NB-CDCCIA, and it causes a decrease in the TDX.

\begin{figure}[htbp]
\begin{center}
\includegraphics[width=0.42\textwidth]{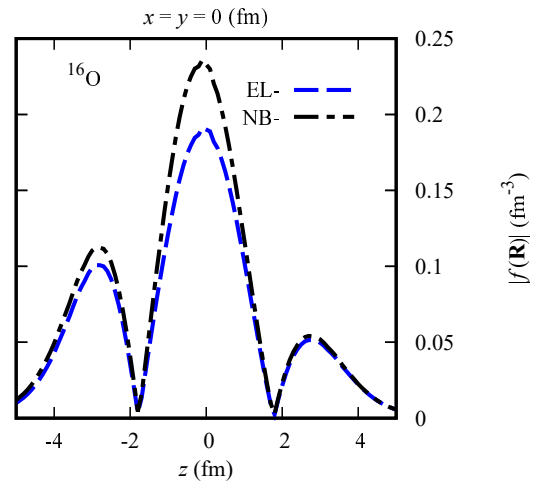}
\caption{Norm of $f (\vR)$ defined by Eq.~\eqref{eq30} at the recoilless condition, $T^\trL_1 = 150$~MeV, for $^{16}$O($p,pd$)$^{14}$N along the direction of the outgoing deuteron; the direction of the deuteron is taken to be parallel to the $z$-axis in this plot. The meaning of the line types is the same as in Fig.~\ref{fig4}(a). \label{fig5}}
\end{center}
\end{figure}

To confirm the above interpretation, the quantity
\begin{align}
f (\vR) \equiv
\frac{1}{\sqrt{4 \pi}}
\varphi_d (R)
\chi^{(\txr{X}) (-)}_{2, \vK_2} (\vR)
\label{eq30}
\end{align}
is defined. The coefficient $1 / \sqrt{4 \pi}$ comes from the spherical harmonics $Y_{00}$ and X = EL or NB. It should be noted that $\chi^{(\txr{EL})}_{2, \vK_2}$ and $\chi^{(\txr{NB})}_{2, \vK_2}$ are the same as $\chi^{(0)}_{2, \vK_2}$ in Eqs.~\eqref{eq7} and \eqref{eq26}, respectively. The EL- and NB-CDCCIA calculations differ only in terms of whether $\chi^{(\txr{X})}_{2, \vK_2}$ includes the back-coupling effect or not, and the other quantities (e.g., $\chi_{0, \vK_0}$ and $\chi_{1, \vK_1}$) are the same. Therefore, we can see the role of the back-coupling effect by comparing $f (\vR)$ of EL- and NB-CDCCIA. Figure~\ref{fig5} shows the norm of $f (\vR)$ at the recoilless condition for the $^{16}$O($p,pd$)$^{14}$N along the direction of the outgoing deuteron. The direction of the knocked-out deuteron is taken to be parallel to the $z$-axis in this plot. The results corresponding to EL-CDCCIA (dashed lines in Fig.~\ref{fig4}) and NB-CDCCIA (dot-dashed lines in Fig.~\ref{fig4}) are denoted by the same line types, respectively. One sees that around $z=0$~fm, $\chi^{(\txr{EL})}_{2, \vK_2}$ has a smaller amplitude than that of $\chi^{(\txr{NB})}_{2, \vK_2}$ as expected. The left peak around $z=-3$~fm is a focus in the contour plot on the $z$-$x$ plane, and it also causes the decrease in the TDX for EL-CDCCIA. The difference in the internal region of the nucleus is, however, mainly responsible for the decrease in the TDX as shown.

\section{Summary and perspective \label{sec4}}
The proton-induced deuteron knockout ($p,pd$) reactions at $250$~MeV on $^{16}$O, $^{40}$Ca, and $^{56}$Ni targets have been investigated in view of the deuteron breakup in the elementary process and deuteron reformation by the FSIs. To achieve this purpose, a new reaction model CDCCIA has been constructed, in which the elementary process of the reaction is described with an impulse picture employing a nucleon-nucleon effective interaction and the reformation process in the final state is treated by using CDCC.

It is found that the approach for the elementary process works to reproduce the experimental data of the $p$-$d$ elastic scattering at forward angles ($\lesssim 90^\circ$) at the energies between $120$ and $250$~MeV. The introduced correction factor $C_\txr{corr}$ is almost energy-independent. Its average value of $0.69$ is used in both $d(p,p)pn$ and ($p,pd$) calculations. The backward-angle ($\gtrsim 150^\circ$) data are also described well by the calculation, though some undershooting is found at $250$~MeV. On the other hand, some undershooting is found at middle angles in which three-nucleon forces play important roles. In addition, there is no experimental data of the $d(p,p)pn$ reaction that can directly be compared with the calculated results. The present approach is considered to be applicable to the $(p,pd)$ reaction studies because (i)~in the present work, the ($p,pd$) kinematics are limited in which the corresponding $p$-$d$ elastic scattering is reproduced well and (ii)~the nucleon-nucleon effective interaction is independent of the internal state of the $pn$ system.

In the calculation of the $^{16}$O($p,pd$)$^{14}$N reaction, the deuteron reformation process decreases the TDX by about $21$~\% around the peak. On the other hand, this process increases the TDXs of the $^{40}$Ca($p,pd$)$^{38}$K and $^{56}$Ni($p,pd$)$^{54}$Co reactions by about $11$ and $14$~\%. This phenomenon is caused by the interference between the elastic and breakup channels of the outgoing deuteron. Although the cross section of the deuteron reformation process in the ($p,pd$) reaction is negligibly small, it significantly changes the TDX of the ($p,pd$) reaction through the interference. The back-coupling effect reduces the magnitude of TDX, in which the knocked-out deuteron is once broken up and then forms deuteron in the end. This reduction can be interpreted qualitatively in terms of the conservation of the flux. It is needed to investigate systematically how the deuteron reformation and back-coupling effect contribute to the knockout cross section, e.g., their dependences on the target mass, incident energy, and reaction kinematics.

For more quantitative discussion regarding the experimental data, it should be necessary to improve the reaction model. Evaluating how well the AMA works in the ($p,pd$) reaction is one of the future studies. The validity of this approximation in the ($p,p\alpha$) reaction has been verified in Ref.~\cite{KazukiYoshida16} but has never been done in the ($p,pd$) reaction. Moreover, in the present study, only the $S$-wave deuteron breakup states are taken into account in both the elementary and deuteron reformation processes. The breakup states with higher angular momenta such as the $P$- and $D$-waves will have an important role in those processes. In addition, the wave function of the $pn$ pair inside a nucleus can be different from that of deuteron in free space. Such $pn$ pair can form a deuteron when it interacts with the incident proton; FSIs will also be responsible for the formation of the deuteron after the $pn$ pair is knocked out as it is. A quantitative description of these processes will be necessary to draw a conclusion on the existence of deuteron in nuclei through the systematic comparison with ($p,pd$) reaction observables. Improvement of the present model for the elementary processes will crucially be important. In future, we will describe the elementary processes of the reaction with an exact three-body model, such as the Faddeev-Alt-Grassberger-Sandhas (FAGS) theory, adopting realistic nuclear forces. We also plan to describe the deuteron and $p$-$n$ breakup states using realistic nuclear forces, e.g., the Reid soft core potential, in the calculation of the $d$-B wave function with CDCC \cite{YIseri91}. We are currently working on the formulation and development of the code for this purpose. Using realistic nuclear forces, however, will increase the number of channels involved in the CDCC calculations and make the calculations much more complicated. It may take some time to obtain numerical results.\\

\section*{ACKNOWLEDGEMENTS}
This work has been supported in part by Grants-in-Aid of the Japan Society for the Promotion of Science (Grants No.~JP20K14475, No.~JP21H00125, and No.~JP21H04975). The computation was carried out with the computer facilities at the Research Center for Nuclear Physics, Osaka University. The authors acknowledge T.~Uesaka, S.~Ishikawa, and S.~Ogawa for fruitful discussions.

\bibliography{Paper_CDCCIA}

\begin{thebibliography}{53}
\expandafter\ifx\csname natexlab\endcsname\relax\def\natexlab#1{#1}\fi
\expandafter\ifx\csname bibnamefont\endcsname\relax
  \def\bibnamefont#1{#1}\fi
\expandafter\ifx\csname bibfnamefont\endcsname\relax
  \def\bibfnamefont#1{#1}\fi
\expandafter\ifx\csname citenamefont\endcsname\relax
  \def\citenamefont#1{#1}\fi
\expandafter\ifx\csname url\endcsname\relax
  \def\url#1{\texttt{#1}}\fi
\expandafter\ifx\csname urlprefix\endcsname\relax\def\urlprefix{URL }\fi
\providecommand{\bibinfo}[2]{#2}
\providecommand{\eprint}[2][]{\url{#2}}

\bibitem[{\citenamefont{Kelly}(World Scientific, Singapore, 2013)}]{JJKelly96}
\bibinfo{author}{\bibfnamefont{J.~J.} \bibnamefont{Kelly}},
  \bibinfo{journal}{in \textit{Advances in Nuclear Physics}, vol. 23, edited by
  J. W. Negele and E. Vogt,} p.~\bibinfo{pages}{75} (\bibinfo{year}{World
  Scientific, Singapore, 2013}).

\bibitem[{\citenamefont{Kramer et~al.}(2001)\citenamefont{Kramer, Blok, and
  Lapik\'{a}s}}]{GJKramer01}
\bibinfo{author}{\bibfnamefont{G.~J.} \bibnamefont{Kramer}},
  \bibinfo{author}{\bibfnamefont{H.~P.} \bibnamefont{Blok}}, \bibnamefont{and}
  \bibinfo{author}{\bibfnamefont{L.}~\bibnamefont{Lapik\'{a}s}},
  \bibinfo{journal}{Nuclear Physics A} \textbf{\bibinfo{volume}{679}},
  \bibinfo{pages}{267} (\bibinfo{year}{2001}), ISSN \bibinfo{issn}{0375-9474}.

\bibitem[{\citenamefont{Wakasa et~al.}(2017)\citenamefont{Wakasa, Ogata, and
  Noro}}]{TWakasa17}
\bibinfo{author}{\bibfnamefont{T.}~\bibnamefont{Wakasa}},
  \bibinfo{author}{\bibfnamefont{K.}~\bibnamefont{Ogata}}, \bibnamefont{and}
  \bibinfo{author}{\bibfnamefont{T.}~\bibnamefont{Noro}},
  \bibinfo{journal}{Prog. Part. Nucl. Phys.} \textbf{\bibinfo{volume}{96}},
  \bibinfo{pages}{32} (\bibinfo{year}{2017}).

\bibitem[{\citenamefont{Bertsch}(World Scientific, Singapore,
  2013)}]{GFBertsch13}
\bibinfo{author}{\bibfnamefont{G.~F.} \bibnamefont{Bertsch}},
  \bibinfo{journal}{in \textit{Fifty Years of Nuclear BCS}, edited by R. A.
  Broglia and V. Zelevinsky}  (\bibinfo{year}{World Scientific, Singapore,
  2013}).

\bibitem[{\citenamefont{Brink and Broglia}(Cambridge University Press,
  Cambridge, 2005)}]{DMBrink05}
\bibinfo{author}{\bibfnamefont{D.~M.} \bibnamefont{Brink}} \bibnamefont{and}
  \bibinfo{author}{\bibfnamefont{R.~A.} \bibnamefont{Broglia}},
  \bibinfo{journal}{\textit{Nuclear Superfluidity: Pairing in Finite Systems}}
  (\bibinfo{year}{Cambridge University Press, Cambridge, 2005}).

\bibitem[{\citenamefont{Egiyan et~al.}(2003)\citenamefont{Egiyan, Dashyan,
  Sargsian, Stepanyan, Weinstein, Adams, Ambrozewicz, Anciant, Anghinolfi,
  Asavapibhop et~al.}}]{KSEgiyan03}
\bibinfo{author}{\bibfnamefont{K.~S.} \bibnamefont{Egiyan}},
  \bibinfo{author}{\bibfnamefont{N.}~\bibnamefont{Dashyan}},
  \bibinfo{author}{\bibfnamefont{M.}~\bibnamefont{Sargsian}},
  \bibinfo{author}{\bibfnamefont{S.}~\bibnamefont{Stepanyan}},
  \bibinfo{author}{\bibfnamefont{L.~B.} \bibnamefont{Weinstein}},
  \bibinfo{author}{\bibfnamefont{G.}~\bibnamefont{Adams}},
  \bibinfo{author}{\bibfnamefont{P.}~\bibnamefont{Ambrozewicz}},
  \bibinfo{author}{\bibfnamefont{E.}~\bibnamefont{Anciant}},
  \bibinfo{author}{\bibfnamefont{M.}~\bibnamefont{Anghinolfi}},
  \bibinfo{author}{\bibfnamefont{B.}~\bibnamefont{Asavapibhop}},
  \bibnamefont{et~al.} (\bibinfo{collaboration}{CLAS Collaboration}),
  \bibinfo{journal}{Phys. Rev. C} \textbf{\bibinfo{volume}{68}},
  \bibinfo{pages}{014313} (\bibinfo{year}{2003}).

\bibitem[{\citenamefont{Egiyan et~al.}(2006)\citenamefont{Egiyan, Dashyan,
  Sargsian, Strikman, Weinstein, Adams, Ambrozewicz, Anghinolfi, Asavapibhop,
  Asryan et~al.}}]{KSEgiyan06}
\bibinfo{author}{\bibfnamefont{K.~S.} \bibnamefont{Egiyan}},
  \bibinfo{author}{\bibfnamefont{N.~B.} \bibnamefont{Dashyan}},
  \bibinfo{author}{\bibfnamefont{M.~M.} \bibnamefont{Sargsian}},
  \bibinfo{author}{\bibfnamefont{M.~I.} \bibnamefont{Strikman}},
  \bibinfo{author}{\bibfnamefont{L.~B.} \bibnamefont{Weinstein}},
  \bibinfo{author}{\bibfnamefont{G.}~\bibnamefont{Adams}},
  \bibinfo{author}{\bibfnamefont{P.}~\bibnamefont{Ambrozewicz}},
  \bibinfo{author}{\bibfnamefont{M.}~\bibnamefont{Anghinolfi}},
  \bibinfo{author}{\bibfnamefont{B.}~\bibnamefont{Asavapibhop}},
  \bibinfo{author}{\bibfnamefont{G.}~\bibnamefont{Asryan}},
  \bibnamefont{et~al.} (\bibinfo{collaboration}{CLAS Collaboration}),
  \bibinfo{journal}{Phys. Rev. Lett.} \textbf{\bibinfo{volume}{96}},
  \bibinfo{pages}{082501} (\bibinfo{year}{2006}).

\bibitem[{\citenamefont{Subedi et~al.}(2008)\citenamefont{Subedi, Shneor,
  Monaghan, Anderson, Aniol, Annand, Arrington, Benaoum, Benmokhtar, Boeglin
  et~al.}}]{RSubedi08}
\bibinfo{author}{\bibfnamefont{R.}~\bibnamefont{Subedi}},
  \bibinfo{author}{\bibfnamefont{R.}~\bibnamefont{Shneor}},
  \bibinfo{author}{\bibfnamefont{P.}~\bibnamefont{Monaghan}},
  \bibinfo{author}{\bibfnamefont{B.~D.} \bibnamefont{Anderson}},
  \bibinfo{author}{\bibfnamefont{K.}~\bibnamefont{Aniol}},
  \bibinfo{author}{\bibfnamefont{J.}~\bibnamefont{Annand}},
  \bibinfo{author}{\bibfnamefont{J.}~\bibnamefont{Arrington}},
  \bibinfo{author}{\bibfnamefont{H.}~\bibnamefont{Benaoum}},
  \bibinfo{author}{\bibfnamefont{F.}~\bibnamefont{Benmokhtar}},
  \bibinfo{author}{\bibfnamefont{W.}~\bibnamefont{Boeglin}},
  \bibnamefont{et~al.}, \bibinfo{journal}{Science}
  \textbf{\bibinfo{volume}{320}}, \bibinfo{pages}{1476} (\bibinfo{year}{2008}).

\bibitem[{\citenamefont{Shneor et~al.}(2007)\citenamefont{Shneor, Monaghan,
  Subedi, Anderson, Aniol, Annand, Arrington, Benaoum, Benmokhtar, Bertin
  et~al.}}]{RShneor07}
\bibinfo{author}{\bibfnamefont{R.}~\bibnamefont{Shneor}},
  \bibinfo{author}{\bibfnamefont{P.}~\bibnamefont{Monaghan}},
  \bibinfo{author}{\bibfnamefont{R.}~\bibnamefont{Subedi}},
  \bibinfo{author}{\bibfnamefont{B.~D.} \bibnamefont{Anderson}},
  \bibinfo{author}{\bibfnamefont{K.}~\bibnamefont{Aniol}},
  \bibinfo{author}{\bibfnamefont{J.}~\bibnamefont{Annand}},
  \bibinfo{author}{\bibfnamefont{J.}~\bibnamefont{Arrington}},
  \bibinfo{author}{\bibfnamefont{H.}~\bibnamefont{Benaoum}},
  \bibinfo{author}{\bibfnamefont{F.}~\bibnamefont{Benmokhtar}},
  \bibinfo{author}{\bibfnamefont{P.}~\bibnamefont{Bertin}},
  \bibnamefont{et~al.} (\bibinfo{collaboration}{Jefferson Lab Hall A
  Collaboration}), \bibinfo{journal}{Phys. Rev. Lett.}
  \textbf{\bibinfo{volume}{99}}, \bibinfo{pages}{072501}
  (\bibinfo{year}{2007}).

\bibitem[{\citenamefont{Sargsian et~al.}(2005)\citenamefont{Sargsian,
  Abrahamyan, Strikman, and Frankfurt}}]{MMSargsian05}
\bibinfo{author}{\bibfnamefont{M.~M.} \bibnamefont{Sargsian}},
  \bibinfo{author}{\bibfnamefont{T.~V.} \bibnamefont{Abrahamyan}},
  \bibinfo{author}{\bibfnamefont{M.~I.} \bibnamefont{Strikman}},
  \bibnamefont{and} \bibinfo{author}{\bibfnamefont{L.~L.}
  \bibnamefont{Frankfurt}}, \bibinfo{journal}{Phys. Rev. C}
  \textbf{\bibinfo{volume}{71}}, \bibinfo{pages}{044615}
  (\bibinfo{year}{2005}).

\bibitem[{\citenamefont{Schiavilla et~al.}(2007)\citenamefont{Schiavilla,
  Wiringa, Pieper, and Carlson}}]{RSchiavilla07}
\bibinfo{author}{\bibfnamefont{R.}~\bibnamefont{Schiavilla}},
  \bibinfo{author}{\bibfnamefont{R.~B.} \bibnamefont{Wiringa}},
  \bibinfo{author}{\bibfnamefont{S.~C.} \bibnamefont{Pieper}},
  \bibnamefont{and} \bibinfo{author}{\bibfnamefont{J.}~\bibnamefont{Carlson}},
  \bibinfo{journal}{Phys. Rev. Lett.} \textbf{\bibinfo{volume}{98}},
  \bibinfo{pages}{132501} (\bibinfo{year}{2007}).

\bibitem[{\citenamefont{Goodman}(1998)}]{ALGoodman98}
\bibinfo{author}{\bibfnamefont{A.~L.} \bibnamefont{Goodman}},
  \bibinfo{journal}{Phys. Rev. C} \textbf{\bibinfo{volume}{58}},
  \bibinfo{pages}{R3051} (\bibinfo{year}{1998}).

\bibitem[{\citenamefont{Frauendorf and Macchiavelli}(2014)}]{SFrauendorf14}
\bibinfo{author}{\bibfnamefont{S.}~\bibnamefont{Frauendorf}} \bibnamefont{and}
  \bibinfo{author}{\bibfnamefont{A.~O.} \bibnamefont{Macchiavelli}},
  \bibinfo{journal}{Prog. Part. Nucl. Phys.} \textbf{\bibinfo{volume}{78}},
  \bibinfo{pages}{24} (\bibinfo{year}{2014}).

\bibitem[{\citenamefont{Yoshida}(2014)}]{KenichiYoshida14}
\bibinfo{author}{\bibfnamefont{K.}~\bibnamefont{Yoshida}},
  \bibinfo{journal}{Phys. Rev. C} \textbf{\bibinfo{volume}{90}},
  \bibinfo{pages}{031303(R)} (\bibinfo{year}{2014}).

\bibitem[{\citenamefont{Litvinova et~al.}(2018)\citenamefont{Litvinova, Robin,
  and Egorova}}]{ELitvinova18}
\bibinfo{author}{\bibfnamefont{E.}~\bibnamefont{Litvinova}},
  \bibinfo{author}{\bibfnamefont{C.}~\bibnamefont{Robin}}, \bibnamefont{and}
  \bibinfo{author}{\bibfnamefont{I.~A.} \bibnamefont{Egorova}},
  \bibinfo{journal}{Phys. Lett. B} \textbf{\bibinfo{volume}{776}},
  \bibinfo{pages}{72} (\bibinfo{year}{2018}).

\bibitem[{\citenamefont{Masui and Kimura}(2016)}]{HMasui16}
\bibinfo{author}{\bibfnamefont{H.}~\bibnamefont{Masui}} \bibnamefont{and}
  \bibinfo{author}{\bibfnamefont{M.}~\bibnamefont{Kimura}},
  \bibinfo{journal}{Prog. Theor. Exp. Phys.} \textbf{\bibinfo{volume}{2016}}
  (\bibinfo{year}{2016}), \bibinfo{note}{053D01}.

\bibitem[{\citenamefont{Samanta et~al.}(1982)\citenamefont{Samanta, Chant,
  Roos, Nadasen, and Cowley}}]{CSamanta82}
\bibinfo{author}{\bibfnamefont{C.}~\bibnamefont{Samanta}},
  \bibinfo{author}{\bibfnamefont{N.~S.} \bibnamefont{Chant}},
  \bibinfo{author}{\bibfnamefont{P.~G.} \bibnamefont{Roos}},
  \bibinfo{author}{\bibfnamefont{A.}~\bibnamefont{Nadasen}}, \bibnamefont{and}
  \bibinfo{author}{\bibfnamefont{A.~A.} \bibnamefont{Cowley}},
  \bibinfo{journal}{Phys. Rev. C} \textbf{\bibinfo{volume}{26}},
  \bibinfo{pages}{1379} (\bibinfo{year}{1982}).

\bibitem[{\citenamefont{Samanta et~al.}(1986)\citenamefont{Samanta, Chant,
  Roos, Nadasen, Wesick, and Cowley}}]{CSamanta86}
\bibinfo{author}{\bibfnamefont{C.}~\bibnamefont{Samanta}},
  \bibinfo{author}{\bibfnamefont{N.~S.} \bibnamefont{Chant}},
  \bibinfo{author}{\bibfnamefont{P.~G.} \bibnamefont{Roos}},
  \bibinfo{author}{\bibfnamefont{A.}~\bibnamefont{Nadasen}},
  \bibinfo{author}{\bibfnamefont{J.}~\bibnamefont{Wesick}}, \bibnamefont{and}
  \bibinfo{author}{\bibfnamefont{A.~A.} \bibnamefont{Cowley}},
  \bibinfo{journal}{Phys. Rev. C} \textbf{\bibinfo{volume}{34}},
  \bibinfo{pages}{1610} (\bibinfo{year}{1986}).

\bibitem[{\citenamefont{Jacob and Maris}(1966)}]{GJacob66}
\bibinfo{author}{\bibfnamefont{G.}~\bibnamefont{Jacob}} \bibnamefont{and}
  \bibinfo{author}{\bibfnamefont{T.~A.~J.} \bibnamefont{Maris}},
  \bibinfo{journal}{Rev. Mod. Phys.} \textbf{\bibinfo{volume}{38}},
  \bibinfo{pages}{121} (\bibinfo{year}{1966}).

\bibitem[{\citenamefont{Jacob and Maris}(1973)}]{GJacob73}
\bibinfo{author}{\bibfnamefont{G.}~\bibnamefont{Jacob}} \bibnamefont{and}
  \bibinfo{author}{\bibfnamefont{T.~A.~J.} \bibnamefont{Maris}},
  \bibinfo{journal}{Rev. Mod. Phys.} \textbf{\bibinfo{volume}{45}},
  \bibinfo{pages}{6} (\bibinfo{year}{1973}).

\bibitem[{\citenamefont{Chant and Roos}(1977)}]{NSChant77}
\bibinfo{author}{\bibfnamefont{N.~S.} \bibnamefont{Chant}} \bibnamefont{and}
  \bibinfo{author}{\bibfnamefont{P.~G.} \bibnamefont{Roos}},
  \bibinfo{journal}{Phys. Rev. C} \textbf{\bibinfo{volume}{15}},
  \bibinfo{pages}{57} (\bibinfo{year}{1977}).

\bibitem[{\citenamefont{Chant and Roos}(1983)}]{NSChant83}
\bibinfo{author}{\bibfnamefont{N.~S.} \bibnamefont{Chant}} \bibnamefont{and}
  \bibinfo{author}{\bibfnamefont{P.~G.} \bibnamefont{Roos}},
  \bibinfo{journal}{Phys. Rev. C} \textbf{\bibinfo{volume}{27}},
  \bibinfo{pages}{1060} (\bibinfo{year}{1983}).

\bibitem[{\citenamefont{Kamimura et~al.}(1986)\citenamefont{Kamimura, Yahiro,
  Iseri, Sakuragi, Kameyama, and Kawai}}]{MKamimura86}
\bibinfo{author}{\bibfnamefont{M.}~\bibnamefont{Kamimura}},
  \bibinfo{author}{\bibfnamefont{M.}~\bibnamefont{Yahiro}},
  \bibinfo{author}{\bibfnamefont{Y.}~\bibnamefont{Iseri}},
  \bibinfo{author}{\bibfnamefont{Y.}~\bibnamefont{Sakuragi}},
  \bibinfo{author}{\bibfnamefont{H.}~\bibnamefont{Kameyama}}, \bibnamefont{and}
  \bibinfo{author}{\bibfnamefont{M.}~\bibnamefont{Kawai}},
  \bibinfo{journal}{Prog. Theor. Phys. Suppl.} \textbf{\bibinfo{volume}{89}},
  \bibinfo{pages}{1} (\bibinfo{year}{1986}).

\bibitem[{\citenamefont{Austern et~al.}(1987)\citenamefont{Austern, Iseri,
  Kamimura, Kawai, Rawitscher, and Yahiro}}]{NAustern87}
\bibinfo{author}{\bibfnamefont{N.}~\bibnamefont{Austern}},
  \bibinfo{author}{\bibfnamefont{Y.}~\bibnamefont{Iseri}},
  \bibinfo{author}{\bibfnamefont{M.}~\bibnamefont{Kamimura}},
  \bibinfo{author}{\bibfnamefont{M.}~\bibnamefont{Kawai}},
  \bibinfo{author}{\bibfnamefont{G.}~\bibnamefont{Rawitscher}},
  \bibnamefont{and} \bibinfo{author}{\bibfnamefont{M.}~\bibnamefont{Yahiro}},
  \bibinfo{journal}{Phys. Rep.} \textbf{\bibinfo{volume}{154}},
  \bibinfo{pages}{125} (\bibinfo{year}{1987}).

\bibitem[{\citenamefont{Yahiro et~al.}(2012)\citenamefont{Yahiro, Ogata,
  Matsumoto, and Minomo}}]{MYahiro12}
\bibinfo{author}{\bibfnamefont{M.}~\bibnamefont{Yahiro}},
  \bibinfo{author}{\bibfnamefont{K.}~\bibnamefont{Ogata}},
  \bibinfo{author}{\bibfnamefont{T.}~\bibnamefont{Matsumoto}},
  \bibnamefont{and} \bibinfo{author}{\bibfnamefont{K.}~\bibnamefont{Minomo}},
  \bibinfo{journal}{Prog. Theor. Exp. Phys.} \textbf{\bibinfo{volume}{2012}}
  (\bibinfo{year}{2012}), \bibinfo{note}{01A206}.

\bibitem[{\citenamefont{Harada and Hirabayashi}(2015)}]{THarada15}
\bibinfo{author}{\bibfnamefont{T.}~\bibnamefont{Harada}} \bibnamefont{and}
  \bibinfo{author}{\bibfnamefont{Y.}~\bibnamefont{Hirabayashi}},
  \bibinfo{journal}{Nucl. Phys. A} \textbf{\bibinfo{volume}{934}},
  \bibinfo{pages}{8} (\bibinfo{year}{2015}), ISSN \bibinfo{issn}{0375-9474}.

\bibitem[{\citenamefont{Yoshida et~al.}(2016)\citenamefont{Yoshida, Minomo, and
  Ogata}}]{KazukiYoshida16}
\bibinfo{author}{\bibfnamefont{K.}~\bibnamefont{Yoshida}},
  \bibinfo{author}{\bibfnamefont{K.}~\bibnamefont{Minomo}}, \bibnamefont{and}
  \bibinfo{author}{\bibfnamefont{K.}~\bibnamefont{Ogata}},
  \bibinfo{journal}{Phys. Rev. C} \textbf{\bibinfo{volume}{94}},
  \bibinfo{pages}{044604} (\bibinfo{year}{2016}).

\bibitem[{\citenamefont{Kudo and Miyazaki}(1986)}]{YKudo86}
\bibinfo{author}{\bibfnamefont{Y.}~\bibnamefont{Kudo}} \bibnamefont{and}
  \bibinfo{author}{\bibfnamefont{K.}~\bibnamefont{Miyazaki}},
  \bibinfo{journal}{Phys. Rev. C} \textbf{\bibinfo{volume}{34}},
  \bibinfo{pages}{1192} (\bibinfo{year}{1986}).

\bibitem[{\citenamefont{Kudo et~al.}(1988)\citenamefont{Kudo, Kanayama, and
  Wakasugi}}]{YKudo88}
\bibinfo{author}{\bibfnamefont{Y.}~\bibnamefont{Kudo}},
  \bibinfo{author}{\bibfnamefont{N.}~\bibnamefont{Kanayama}}, \bibnamefont{and}
  \bibinfo{author}{\bibfnamefont{T.}~\bibnamefont{Wakasugi}},
  \bibinfo{journal}{Phys. Rev. C} \textbf{\bibinfo{volume}{38}},
  \bibinfo{pages}{1126} (\bibinfo{year}{1988}).

\bibitem[{\citenamefont{Kanayama et~al.}(1990)\citenamefont{Kanayama, Kudo,
  Tsunoda, and Wakasugi}}]{NKanayama90}
\bibinfo{author}{\bibfnamefont{N.}~\bibnamefont{Kanayama}},
  \bibinfo{author}{\bibfnamefont{Y.}~\bibnamefont{Kudo}},
  \bibinfo{author}{\bibfnamefont{H.}~\bibnamefont{Tsunoda}}, \bibnamefont{and}
  \bibinfo{author}{\bibfnamefont{T.}~\bibnamefont{Wakasugi}},
  \bibinfo{journal}{Prog. Theor. Phys.} \textbf{\bibinfo{volume}{83}},
  \bibinfo{pages}{540} (\bibinfo{year}{1990}).

\bibitem[{\citenamefont{Kudo and Tsunoda}(1993)}]{YKudo93}
\bibinfo{author}{\bibfnamefont{Y.}~\bibnamefont{Kudo}} \bibnamefont{and}
  \bibinfo{author}{\bibfnamefont{H.}~\bibnamefont{Tsunoda}},
  \bibinfo{journal}{Prog. Theor. Phys.} \textbf{\bibinfo{volume}{89}},
  \bibinfo{pages}{89} (\bibinfo{year}{1993}), ISSN \bibinfo{issn}{0033-068X}.

\bibitem[{\citenamefont{Ogata et~al.}(2015)\citenamefont{Ogata, Yoshida, and
  Minomo}}]{KOgata15}
\bibinfo{author}{\bibfnamefont{K.}~\bibnamefont{Ogata}},
  \bibinfo{author}{\bibfnamefont{K.}~\bibnamefont{Yoshida}}, \bibnamefont{and}
  \bibinfo{author}{\bibfnamefont{K.}~\bibnamefont{Minomo}},
  \bibinfo{journal}{Phys. Rev. C} \textbf{\bibinfo{volume}{92}},
  \bibinfo{pages}{034616} (\bibinfo{year}{2015}).

\bibitem[{\citenamefont{Chazono}(2022)}]{YChazono22}
\bibinfo{author}{\bibfnamefont{Y.}~\bibnamefont{Chazono}},
  \bibinfo{journal}{Ph.D. thesis, Osaka University}  (\bibinfo{year}{2022}),
  \urlprefix\url{https://doi.org/10.18910/87811}.

\bibitem[{\citenamefont{Franey and Love}(1985)}]{MAFraney85}
\bibinfo{author}{\bibfnamefont{M.~A.} \bibnamefont{Franey}} \bibnamefont{and}
  \bibinfo{author}{\bibfnamefont{W.~G.} \bibnamefont{Love}},
  \bibinfo{journal}{Phys. Rev. C} \textbf{\bibinfo{volume}{31}},
  \bibinfo{pages}{488} (\bibinfo{year}{1985}).

\bibitem[{\citenamefont{Matsumoto et~al.}(2003)\citenamefont{Matsumoto,
  Kamizato, Ogata, Iseri, Hiyama, Kamimura, and Yahiro}}]{TMatsumoto03}
\bibinfo{author}{\bibfnamefont{T.}~\bibnamefont{Matsumoto}},
  \bibinfo{author}{\bibfnamefont{T.}~\bibnamefont{Kamizato}},
  \bibinfo{author}{\bibfnamefont{K.}~\bibnamefont{Ogata}},
  \bibinfo{author}{\bibfnamefont{Y.}~\bibnamefont{Iseri}},
  \bibinfo{author}{\bibfnamefont{E.}~\bibnamefont{Hiyama}},
  \bibinfo{author}{\bibfnamefont{M.}~\bibnamefont{Kamimura}}, \bibnamefont{and}
  \bibinfo{author}{\bibfnamefont{M.}~\bibnamefont{Yahiro}},
  \bibinfo{journal}{Phys. Rev. C} \textbf{\bibinfo{volume}{68}},
  \bibinfo{pages}{064607} (\bibinfo{year}{2003}).

\bibitem[{\citenamefont{Ohmura et~al.}(1970)\citenamefont{Ohmura, Imanishi,
  Ichimura, and Kawai}}]{TOhmura70}
\bibinfo{author}{\bibfnamefont{T.}~\bibnamefont{Ohmura}},
  \bibinfo{author}{\bibfnamefont{B.}~\bibnamefont{Imanishi}},
  \bibinfo{author}{\bibfnamefont{M.}~\bibnamefont{Ichimura}}, \bibnamefont{and}
  \bibinfo{author}{\bibfnamefont{M.}~\bibnamefont{Kawai}},
  \bibinfo{journal}{Prog. Theor. Phys.} \textbf{\bibinfo{volume}{43}},
  \bibinfo{pages}{347} (\bibinfo{year}{1970}).

\bibitem[{\citenamefont{Ermisch et~al.}(2005)\citenamefont{Ermisch,
  Amir-Ahmadi, van~den Berg, Castelijns, Davids, Deltuva, Epelbaum, Gl\"ockle,
  Golak, Harakeh et~al.}}]{KErmisch05}
\bibinfo{author}{\bibfnamefont{K.}~\bibnamefont{Ermisch}},
  \bibinfo{author}{\bibfnamefont{H.~R.} \bibnamefont{Amir-Ahmadi}},
  \bibinfo{author}{\bibfnamefont{A.~M.} \bibnamefont{van~den Berg}},
  \bibinfo{author}{\bibfnamefont{R.}~\bibnamefont{Castelijns}},
  \bibinfo{author}{\bibfnamefont{B.}~\bibnamefont{Davids}},
  \bibinfo{author}{\bibfnamefont{A.}~\bibnamefont{Deltuva}},
  \bibinfo{author}{\bibfnamefont{E.}~\bibnamefont{Epelbaum}},
  \bibinfo{author}{\bibfnamefont{W.}~\bibnamefont{Gl\"ockle}},
  \bibinfo{author}{\bibfnamefont{J.}~\bibnamefont{Golak}},
  \bibinfo{author}{\bibfnamefont{M.~N.} \bibnamefont{Harakeh}},
  \bibnamefont{et~al.}, \bibinfo{journal}{Phys. Rev. C}
  \textbf{\bibinfo{volume}{71}}, \bibinfo{pages}{064004}
  (\bibinfo{year}{2005}).

\bibitem[{\citenamefont{Kuroda et~al.}(1964)\citenamefont{Kuroda, Michalowicz,
  and Poulet}}]{KKuroda64}
\bibinfo{author}{\bibfnamefont{K.}~\bibnamefont{Kuroda}},
  \bibinfo{author}{\bibfnamefont{A.}~\bibnamefont{Michalowicz}},
  \bibnamefont{and} \bibinfo{author}{\bibfnamefont{M.}~\bibnamefont{Poulet}},
  \bibinfo{journal}{Phys. Lett.} \textbf{\bibinfo{volume}{13}},
  \bibinfo{pages}{67} (\bibinfo{year}{1964}).

\bibitem[{\citenamefont{Hatanaka et~al.}(2002)\citenamefont{Hatanaka, Shimizu,
  Hirooka, Kamiya, Kitamura, Maeda, Noro, Obayashi, Sagara, Saito
  et~al.}}]{KHatanaka02}
\bibinfo{author}{\bibfnamefont{K.}~\bibnamefont{Hatanaka}},
  \bibinfo{author}{\bibfnamefont{Y.}~\bibnamefont{Shimizu}},
  \bibinfo{author}{\bibfnamefont{D.}~\bibnamefont{Hirooka}},
  \bibinfo{author}{\bibfnamefont{J.}~\bibnamefont{Kamiya}},
  \bibinfo{author}{\bibfnamefont{Y.}~\bibnamefont{Kitamura}},
  \bibinfo{author}{\bibfnamefont{Y.}~\bibnamefont{Maeda}},
  \bibinfo{author}{\bibfnamefont{T.}~\bibnamefont{Noro}},
  \bibinfo{author}{\bibfnamefont{E.}~\bibnamefont{Obayashi}},
  \bibinfo{author}{\bibfnamefont{K.}~\bibnamefont{Sagara}},
  \bibinfo{author}{\bibfnamefont{T.}~\bibnamefont{Saito}},
  \bibnamefont{et~al.}, \bibinfo{journal}{Phys. Rev. C}
  \textbf{\bibinfo{volume}{66}}, \bibinfo{pages}{044002}
  (\bibinfo{year}{2002}).

\bibitem[{\citenamefont{Terashima et~al.}(2018)\citenamefont{Terashima, Yu,
  Ong, Tanihata, Adachi, Aoi, Chan, Fujioka, Fukuda, Geissel
  et~al.}}]{STerashima18}
\bibinfo{author}{\bibfnamefont{S.}~\bibnamefont{Terashima}},
  \bibinfo{author}{\bibfnamefont{L.}~\bibnamefont{Yu}},
  \bibinfo{author}{\bibfnamefont{H.~J.} \bibnamefont{Ong}},
  \bibinfo{author}{\bibfnamefont{I.}~\bibnamefont{Tanihata}},
  \bibinfo{author}{\bibfnamefont{S.}~\bibnamefont{Adachi}},
  \bibinfo{author}{\bibfnamefont{N.}~\bibnamefont{Aoi}},
  \bibinfo{author}{\bibfnamefont{P.~Y.} \bibnamefont{Chan}},
  \bibinfo{author}{\bibfnamefont{H.}~\bibnamefont{Fujioka}},
  \bibinfo{author}{\bibfnamefont{M.}~\bibnamefont{Fukuda}},
  \bibinfo{author}{\bibfnamefont{H.}~\bibnamefont{Geissel}},
  \bibnamefont{et~al.}, \bibinfo{journal}{Phys. Rev. Lett.}
  \textbf{\bibinfo{volume}{121}}, \bibinfo{pages}{242501}
  (\bibinfo{year}{2018}).

\bibitem[{\citenamefont{Wita\l{}a et~al.}(1998)\citenamefont{Wita\l{}a,
  Gl\"ockle, H\"uber, Golak, and Kamada}}]{HWitala98}
\bibinfo{author}{\bibfnamefont{H.}~\bibnamefont{Wita\l{}a}},
  \bibinfo{author}{\bibfnamefont{W.}~\bibnamefont{Gl\"ockle}},
  \bibinfo{author}{\bibfnamefont{D.}~\bibnamefont{H\"uber}},
  \bibinfo{author}{\bibfnamefont{J.}~\bibnamefont{Golak}}, \bibnamefont{and}
  \bibinfo{author}{\bibfnamefont{H.}~\bibnamefont{Kamada}},
  \bibinfo{journal}{Phys. Rev. Lett.} \textbf{\bibinfo{volume}{81}},
  \bibinfo{pages}{1183} (\bibinfo{year}{1998}).

\bibitem[{\citenamefont{Nemoto et~al.}(1998)\citenamefont{Nemoto, Chmielewski,
  Oryu, and Sauer}}]{SNemoto98}
\bibinfo{author}{\bibfnamefont{S.}~\bibnamefont{Nemoto}},
  \bibinfo{author}{\bibfnamefont{K.}~\bibnamefont{Chmielewski}},
  \bibinfo{author}{\bibfnamefont{S.}~\bibnamefont{Oryu}}, \bibnamefont{and}
  \bibinfo{author}{\bibfnamefont{P.~U.} \bibnamefont{Sauer}},
  \bibinfo{journal}{Phys. Rev. C} \textbf{\bibinfo{volume}{58}},
  \bibinfo{pages}{2599} (\bibinfo{year}{1998}).

\bibitem[{\citenamefont{Abfalterer et~al.}(1998)\citenamefont{Abfalterer,
  Bateman, Dietrich, Elster, Finlay, Gl\"ockle, Golak, Haight, H\"uber, Morgan
  et~al.}}]{WPAbfalterer98}
\bibinfo{author}{\bibfnamefont{W.~P.} \bibnamefont{Abfalterer}},
  \bibinfo{author}{\bibfnamefont{F.~B.} \bibnamefont{Bateman}},
  \bibinfo{author}{\bibfnamefont{F.~S.} \bibnamefont{Dietrich}},
  \bibinfo{author}{\bibfnamefont{C.}~\bibnamefont{Elster}},
  \bibinfo{author}{\bibfnamefont{R.~W.} \bibnamefont{Finlay}},
  \bibinfo{author}{\bibfnamefont{W.}~\bibnamefont{Gl\"ockle}},
  \bibinfo{author}{\bibfnamefont{J.}~\bibnamefont{Golak}},
  \bibinfo{author}{\bibfnamefont{R.~C.} \bibnamefont{Haight}},
  \bibinfo{author}{\bibfnamefont{D.}~\bibnamefont{H\"uber}},
  \bibinfo{author}{\bibfnamefont{G.~L.} \bibnamefont{Morgan}},
  \bibnamefont{et~al.}, \bibinfo{journal}{Phys. Rev. Lett.}
  \textbf{\bibinfo{volume}{81}}, \bibinfo{pages}{57} (\bibinfo{year}{1998}).

\bibitem[{\citenamefont{Wita\l{}a et~al.}(1999)\citenamefont{Wita\l{}a, Kamada,
  Nogga, Gl\"ockle, Elster, and H\"uber}}]{HWitala99}
\bibinfo{author}{\bibfnamefont{H.}~\bibnamefont{Wita\l{}a}},
  \bibinfo{author}{\bibfnamefont{H.}~\bibnamefont{Kamada}},
  \bibinfo{author}{\bibfnamefont{A.}~\bibnamefont{Nogga}},
  \bibinfo{author}{\bibfnamefont{W.}~\bibnamefont{Gl\"ockle}},
  \bibinfo{author}{\bibfnamefont{C.}~\bibnamefont{Elster}}, \bibnamefont{and}
  \bibinfo{author}{\bibfnamefont{D.}~\bibnamefont{H\"uber}},
  \bibinfo{journal}{Phys. Rev. C} \textbf{\bibinfo{volume}{59}},
  \bibinfo{pages}{3035} (\bibinfo{year}{1999}).

\bibitem[{\citenamefont{Sekiguchi et~al.}(2002)\citenamefont{Sekiguchi, Sakai,
  Wita\l{}a, Gl\"ockle, Golak, Hatano, Kamada, Kato, Maeda, Nishikawa
  et~al.}}]{KSekiguchi02}
\bibinfo{author}{\bibfnamefont{K.}~\bibnamefont{Sekiguchi}},
  \bibinfo{author}{\bibfnamefont{H.}~\bibnamefont{Sakai}},
  \bibinfo{author}{\bibfnamefont{H.}~\bibnamefont{Wita\l{}a}},
  \bibinfo{author}{\bibfnamefont{W.}~\bibnamefont{Gl\"ockle}},
  \bibinfo{author}{\bibfnamefont{J.}~\bibnamefont{Golak}},
  \bibinfo{author}{\bibfnamefont{M.}~\bibnamefont{Hatano}},
  \bibinfo{author}{\bibfnamefont{H.}~\bibnamefont{Kamada}},
  \bibinfo{author}{\bibfnamefont{H.}~\bibnamefont{Kato}},
  \bibinfo{author}{\bibfnamefont{Y.}~\bibnamefont{Maeda}},
  \bibinfo{author}{\bibfnamefont{J.}~\bibnamefont{Nishikawa}},
  \bibnamefont{et~al.}, \bibinfo{journal}{Phys. Rev. C}
  \textbf{\bibinfo{volume}{65}}, \bibinfo{pages}{034003}
  (\bibinfo{year}{2002}).

\bibitem[{\citenamefont{Hama et~al.}(1990)\citenamefont{Hama, Clark, Cooper,
  Sherif, and Mercer}}]{SHama90}
\bibinfo{author}{\bibfnamefont{S.}~\bibnamefont{Hama}},
  \bibinfo{author}{\bibfnamefont{B.~C.} \bibnamefont{Clark}},
  \bibinfo{author}{\bibfnamefont{E.~D.} \bibnamefont{Cooper}},
  \bibinfo{author}{\bibfnamefont{H.~S.} \bibnamefont{Sherif}},
  \bibnamefont{and} \bibinfo{author}{\bibfnamefont{R.~L.}
  \bibnamefont{Mercer}}, \bibinfo{journal}{Phys. Rev. C}
  \textbf{\bibinfo{volume}{41}}, \bibinfo{pages}{2737} (\bibinfo{year}{1990}).

\bibitem[{\citenamefont{Cooper et~al.}(1993)\citenamefont{Cooper, Hama, Clark,
  and Mercer}}]{EDCooper93}
\bibinfo{author}{\bibfnamefont{E.~D.} \bibnamefont{Cooper}},
  \bibinfo{author}{\bibfnamefont{S.}~\bibnamefont{Hama}},
  \bibinfo{author}{\bibfnamefont{B.~C.} \bibnamefont{Clark}}, \bibnamefont{and}
  \bibinfo{author}{\bibfnamefont{R.~L.} \bibnamefont{Mercer}},
  \bibinfo{journal}{Phys. Rev. C} \textbf{\bibinfo{volume}{47}},
  \bibinfo{pages}{297} (\bibinfo{year}{1993}).

\bibitem[{\citenamefont{Cooper et~al.}(2009)\citenamefont{Cooper, Hama, and
  Clark}}]{EDCooper09}
\bibinfo{author}{\bibfnamefont{E.~D.} \bibnamefont{Cooper}},
  \bibinfo{author}{\bibfnamefont{S.}~\bibnamefont{Hama}}, \bibnamefont{and}
  \bibinfo{author}{\bibfnamefont{B.~C.} \bibnamefont{Clark}},
  \bibinfo{journal}{Phys. Rev. C} \textbf{\bibinfo{volume}{80}},
  \bibinfo{pages}{034605} (\bibinfo{year}{2009}).

\bibitem[{\citenamefont{Arnold et~al.}(1981)\citenamefont{Arnold, Clark,
  Mercer, and Schwandt}}]{LGArnold81}
\bibinfo{author}{\bibfnamefont{L.~G.} \bibnamefont{Arnold}},
  \bibinfo{author}{\bibfnamefont{B.~C.} \bibnamefont{Clark}},
  \bibinfo{author}{\bibfnamefont{R.~L.} \bibnamefont{Mercer}},
  \bibnamefont{and} \bibinfo{author}{\bibfnamefont{P.}~\bibnamefont{Schwandt}},
  \bibinfo{journal}{Phys. Rev. C} \textbf{\bibinfo{volume}{23}},
  \bibinfo{pages}{1949} (\bibinfo{year}{1981}).

\bibitem[{\citenamefont{Timofeyuk and
  Johnson}(2013{\natexlab{a}})}]{NKTimofeyuk13a}
\bibinfo{author}{\bibfnamefont{N.~K.} \bibnamefont{Timofeyuk}}
  \bibnamefont{and} \bibinfo{author}{\bibfnamefont{R.~C.}
  \bibnamefont{Johnson}}, \bibinfo{journal}{Phys. Rev. Lett.}
  \textbf{\bibinfo{volume}{110}}, \bibinfo{pages}{112501}
  (\bibinfo{year}{2013}{\natexlab{a}}).

\bibitem[{\citenamefont{Timofeyuk and
  Johnson}(2013{\natexlab{b}})}]{NKTimofeyuk13b}
\bibinfo{author}{\bibfnamefont{N.~K.} \bibnamefont{Timofeyuk}}
  \bibnamefont{and} \bibinfo{author}{\bibfnamefont{R.~C.}
  \bibnamefont{Johnson}}, \bibinfo{journal}{Phys. Rev. C}
  \textbf{\bibinfo{volume}{87}}, \bibinfo{pages}{064610}
  (\bibinfo{year}{2013}{\natexlab{b}}).

\bibitem[{\citenamefont{Johnson and Timofeyuk}(2014)}]{RCJohnson14}
\bibinfo{author}{\bibfnamefont{R.~C.} \bibnamefont{Johnson}} \bibnamefont{and}
  \bibinfo{author}{\bibfnamefont{N.~K.} \bibnamefont{Timofeyuk}},
  \bibinfo{journal}{Phys. Rev. C} \textbf{\bibinfo{volume}{89}},
  \bibinfo{pages}{024605} (\bibinfo{year}{2014}).

\bibitem[{\citenamefont{Iseri et~al.}(1991)\citenamefont{Iseri, Tanifuji, Aoki,
  and Kawai}}]{YIseri91}
\bibinfo{author}{\bibfnamefont{Y.}~\bibnamefont{Iseri}},
  \bibinfo{author}{\bibfnamefont{M.}~\bibnamefont{Tanifuji}},
  \bibinfo{author}{\bibfnamefont{Y.}~\bibnamefont{Aoki}}, \bibnamefont{and}
  \bibinfo{author}{\bibfnamefont{M.}~\bibnamefont{Kawai}},
  \bibinfo{journal}{Phys. Lett. B} \textbf{\bibinfo{volume}{265}},
  \bibinfo{pages}{207} (\bibinfo{year}{1991}), ISSN \bibinfo{issn}{0370-2693}.

\end{thebibliography}

\end{document}